\pdfoutput=1
\documentclass[twoside,onecolumn]{article}

\usepackage[english]{babel} 

\usepackage[hmarginratio=1:1,top=32mm,columnsep=20pt]{geometry} 
\usepackage[hang, small,labelfont=bf,up,textfont=it,up]{caption} 

\usepackage{abstract} 

\usepackage{fancyhdr} 
\usepackage{titling} 
\usepackage{graphicx}

\pretitle{\begin{center}\Huge\bfseries} 
\posttitle{\end{center}} 
\title{Robotic Telescopes in Education} 
\author{%
\textsc{E.L. Gomez}\thanks{$^\ast$Corresponding author. Email: egomez@lco.global. Both authors contributed equally to this paper.} \\[1ex] 
\normalsize Las Cumbres Observatory, Cardiff University, UK; \\ 
\normalsize egomez@lco.global 
\and 
\textsc{M.T. Fitzgerald} \\[1ex] 
\normalsize Edith Cowan Institute for Education Research, Perth, Australia \\ 
}
\renewcommand{\maketitlehookd}{%
\begin{abstract}
The power of robotic telescopes to transform science education has been voiced by multiple sources, since the 1980s. Since then, much technical progress has been made in robotic telescope provision to end users via a variety of different approaches. The educational transformation hoped for by the provision of this technology has, so far, yet to be achieved on a scale matching the technical advancements. In this paper, the history, definition, role and rationale of optical robotic telescopes with a focus on their use in education is provided. The current telescope access providers and educational projects and their broad uses in traditional schooling, undergraduate and outreach are then outlined. From this background, the current challenges to the field, which are numerous, are then presented. This review is concluded with a series of recommendations for current and future projects that are apparent and have emerged from the literature.
\end{abstract}
}
\date{Submitted to Astronomical Review: 25 Oct 2016}

\begin{document}

\maketitle


\section{Introduction}

Observations are the fundamental measurements of astronomy as a science. These are matched to data and theories produced by laboratory astrophysicists and other physical scientists to test provable predictions. Astronomers are in a unique position among scientists as they are unable to undertake experiments directly on the subjects of their studies. As astronomers we must wait for photons (and now other forms of non-electromagnetic radiation \cite{GW150914}) to travel through the Universe, towards the Earth and its environment. Key to making discoveries is having an appropriate telescope in an appropriate place to witness these photons and the story they tell.

For most of human history, astronomical observations beyond that which can be seen with the naked eye or with small telescopes or binoculars have been inaccessible except to those involved with scientific research or the relatively few who have the large resources required to run a personal observatory. Even when a telescope may be available, the most suitable observing locations are not likely to be within a reasonable distance of the typical city-dweller. A variety of technologies, most notably CCD cameras, robotic mounts and more efficient, powerful, and affordable computers in the latter half of the 20th Century have provided potential solutions to these issues \cite{Marschall1996, Sadleretal2000} allowing access to a wider public.

From a keen astronomers' perspective, this increasingly better access to higher quality data from better telescopes could in itself seem to be a boon for education. To those outside of the field, expenditure on such access, especially when sourced from public funds, needs to be provided with a motivation. The promise of robotic telescopes for education is usually couched in terms of its capacity to deal with certain local, national or international problems related to STEM. In particular, the distinct lack of interest of students in science as formal schooling goes on \cite{Osborneetal2003} and the 'leaky pipeline' \cite{Cannadyetal2014} is seen as a problem that astronomy, with its aesthetic and general appeal, should be able to address.

In this paper, we will focus on the wide role of robotic telescopes in formal, informal and tertiary education, the challenges in using them, and their future potential. We endeavour to provide a relatively concise description of the history of the field and current state of the field and a discussion of the important issues that need to be addressed for the future.

\subsection{Brief Technical History}\label{class}

A brief technical history of robotic autonomous observatories is provided by CastroTirado \cite{CastroTirado2010} from the earliest attempts in the late 1960s with the University of Wisconsin Telescope up through the emerging computer and technical revolution \cite{Genet1982, Genet1989} until near the present day. An earlier paper by Baruch \cite{Baruch1992} delves into more detail about the early development of robotic astronomy and the interaction between engineering realities and astronomical research.

Many modern observatories have `roboticised' their observing away from electro-mechanical controls to providing software interfaces to drive the mechanical elements. This is not a trivial step. Retrofitting older observatories to become robotic is often prohibitively expensive, but once an observatory can be controlled by software, it paves the way for remote operation and a stronger possibility of using telescopes for education.

Since the early 1990’s, students have been able to take advantage of remotely operated astronomical observatories. These included such early projects as the 14" Remote Access Astronomy Project (RAAP, Figure 1L) at UC Santa Barbara, the original 12" Bradford Robotic Telescope (BRT) \cite{Baruch2000}, five 6" telescopes at MicroObservatory \cite{Sadleretal2000} and a variety of telescopes, including a 30" at UC Berkeley \cite{AsbellClarkeetal1996}, through the Hands-On Universe (HoU) project \cite{Pennypacker1998}. The Telescopes in Education (TIE) project [\cite{Clark1998}, Figure 1R] was among the first to bring observing with a professional observatory (in this case, the 24" Mount Wilson telescope) into the classroom.

\begin{figure}
\includegraphics[width=14cm]{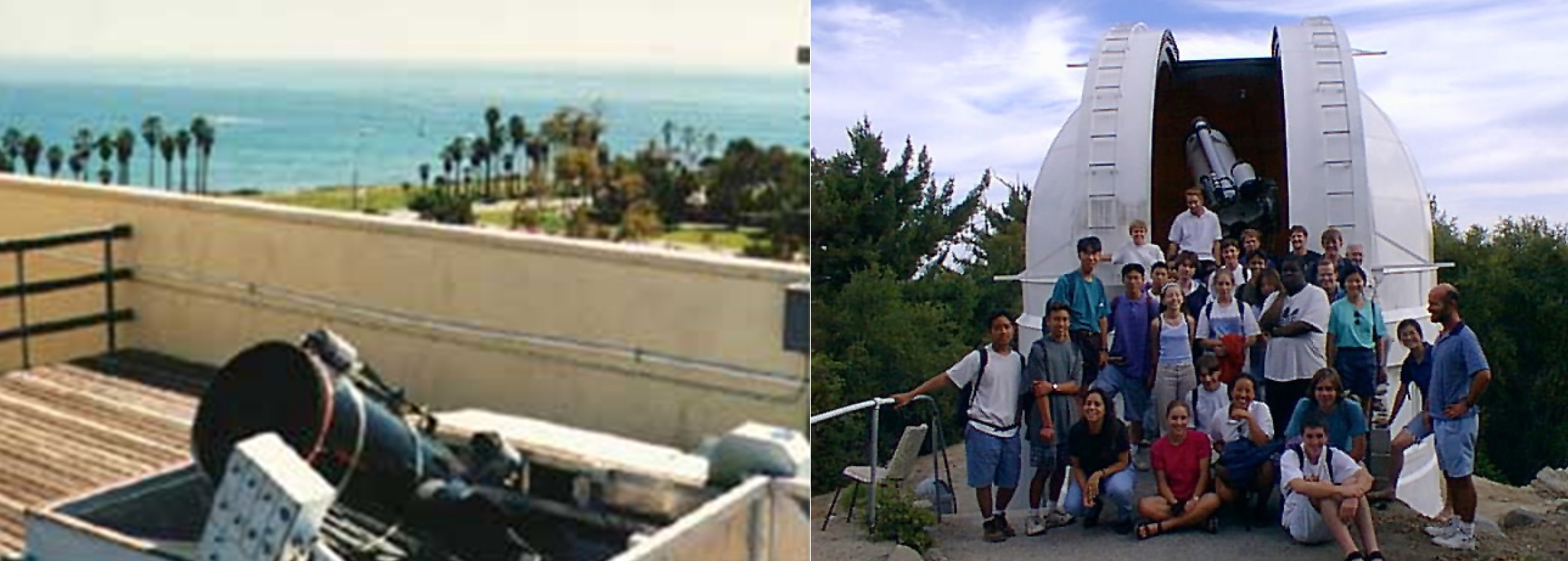}
\caption{Examples of early telescope education projects. The left-hand figure (L) shows the 14" RAAP telescope at UC Santa Barbara. The right-hand figure (R) shows the 24" telescope at Mt. Wilson used by the Telescopes in Education project.}
\label{sample-figure}
\end{figure}

During a similar time period, commercially available, amateur-grade telescopes of 8" to 16" apertures developed the ability to be remote controlled via a simple control pad. This control pad was readily replaced by computer, as standardised control software became available (e.g. ACP from DC-3 Dreams, The Sky from Software Bisque or MPO Connections from BDW Publishing) which provided off-the-shelf ways to roboticise existing equipment. The rise of the Internet of Things has made operating a physical device via the internet a regular, almost trivial, occurrence. Telescopes are no exception, with the majority of commercial, amateur telescopes being able to be remote controlled over great distances, via the internet. It became a logical step to replace the serial connection between telescope and computer with an internet connection for more flexibility. Many universities and schools invested in similar observatories, which needed minimal commissioning and maintenance, but suited their range of pedagogical activities.

\subsection{Remote Control vs Autonomous Control}

For the purpose of this article we define \textit{robotic} as, “being remotely operable”. We refer to instances where a telescope observes under direct human control as \textit{remote control} and instances where a telescope observes without human intervention as \textit{autonomous control}. Many robotic observatories incorporate elements of both \textit{remote control} and \textit{autonomous control}.

Remote Control has been used in two main situations. Having a telescope operator to manage the observing experience paired with a teacher who provides a motivating and educational experience to remote student users \cite{Mckinnnon2005, Clark1998, Lubin1992c}. The second is to facilitate access for research astronomers to access, typically, large aperture telescopes that require manual control at high altitude and distant locations \cite{Zijlstraetal1997}. Both are dependant on the requirement for a fast data transfer system between the observatory and the user \cite{QuerciQuerci2000}.

Robotic observatories open up many more possibilities for educational use. They can be located in relatively hostile or unpleasant locations such as the Atacama desert, Antarctica \cite{Ashleyetal2004}, or even in space, but their robotic nature means that their users can be located anywhere. This gives them a much broader appeal, where the focus can be on the science (and ultimately the education) and not the logistics of operating the telescope.

Networks of telescopes linked together over the internet available for education have been promised by many groups of astronomers over the past few decades \cite{GeldermanPasaIAE2008}. For a long time, there was little stable success with many networks existing for either short periods of time or not at all. Since the mid-late 2000s a variety of pro-amateur institutions (such as iTelescope, Slooh, Sierra Stars Observatory Network) have created and demonstrated medium-term stability in providing robotic telescopes which provide time for education use. In a similar time frame professional research observatories have come online which provide time for specifically education-related purposes (e.g. Las Cumbres Observatory, SkyNET).

\subsection{Problems robotic telescopes solve}\label{class}

For the purposes of education, robotic telescopes solve a number of important problems. School children are at school during the day but most astronomical phenomena are visible during the night. There exist many logistical difficulties to getting school children to an observatory after the school day, as Percy \cite{Percy2003} states the problem: ``the stars come out at night, the students don't".

Access to sites where the sky is not effected by light pollution  is also a problem.  A major issue facing optical astronomers is the increase in light pollution \cite{Kybaetal2015}, especially as old dimmer sodium lamps are being replaced by high power LEDs.  Schools are generally located in the center of population distributions which also are correlated with higher light pollution, as shown in Figure 2. Unless the school is in a rural or remote town, it is likely that most students will be located in a heavily light polluted location.

\begin{figure}
\includegraphics[width=14cm]{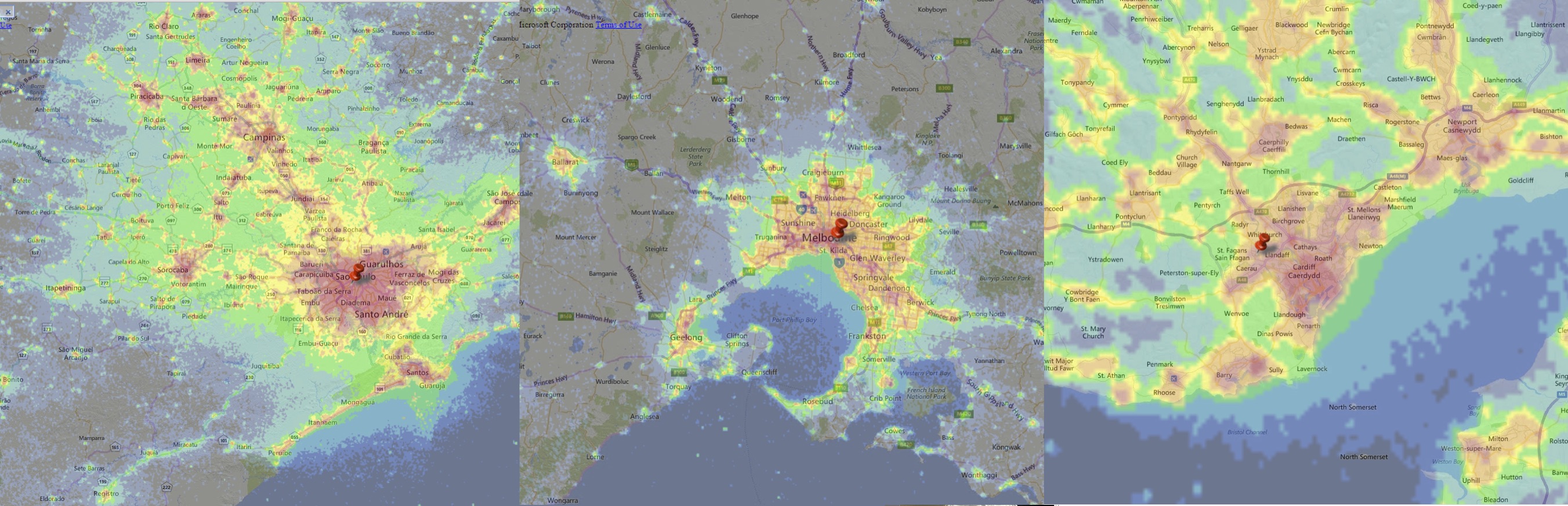}
\caption{Estimates of the light pollution for three typical modern cities, Sao Paulo, Melbourne and Cardiff. The light pollution is highly correlated to the local student population. This image was created using https://www.lightpollutionmap.info/}
\label{sample-figure}
\end{figure}

Even if a suitable dark sky spot were available to a school and an observing night on a cloudless evening could be organised, the quality of the instrumentation available to a typical school would be limited. There are many cases of schools buying or being gifted 4"-16" equatorial telescopes (e.g. \cite{ag2009}) which are either permanently housed on the school premises or available through the science department. For the non-amateur astronomer, setting up such a telescope is a challenge. Equatorial mounted telescopes are much more difficult to set up and align than Dobsonians and as a consequence are often underused.

Even in the highest GDP economies around the world, school budgets are typically very tight and have little room for purchase of such expensive equipment. Only the most well resourced, or philanthropically endowed, schools can afford such expenditure. By providing access to robotic and remote telescopes for free or very low cost, schools in less developed countries and/or in lower socioeconomic areas have equal access to this instrumentation \cite{Sadleretal2001}. Access to a telescope via computer also allows access to those students who, due to physical reasons, may not be able to access ladders or awkward positioned eyepieces. It also allows the visually impaired to use computer accessibility tools to be able to see images from a telescope that they may not be able to through the traditional eyepiece.

Even if a telescope can be configured, a dark sky and a clear night found, and children are able to use this telescope during the night, managing their expectations can be difficult. Viewing objects such as Saturn and the Moon can provide inspiring experiences, but some of the more interesting deep sky objects can be very faint, lacking in colour, and not matching students expectations particularly compared to images found online \cite{Sadleretal2000}.

To truly explore the night sky in \textit{colour}, the eye is insufficient and a more sensitive detector, such as a DSLR camera, is needed. To allow for long exposures the costs increase significantly as adding a good quality tracking mount and high quality optical tube is necessary. Such an investment of time and money would only be available to a vanishingly small fraction of schools. There is no known systematic study of the usage of telescopes of this calibre in schools but informal conversations suggest most are largely unused in long-term storage.

\subsection{The Role of Robotics Telescopes}

By having remote or robotic telescopes accurately and automatically aligned, all of the stress of setting up the telescope and pointing, which requires a lot of skill and practice, is removed from the user. These telescopes can be permanently located at a given spot, meaning that their initial setup only needs to be undertaken once by a small group of professionals who only occasionally need to maintain the telescopes.

There are a variety of benefits to the telescope user. As the telescopes do not need to be transported, more effective mounts and larger apertures can be provided. Physically damaging the telescopes can be made next to impossible by design. There is no time expenditure to setup, pack up or maintain the instrument on the part of the user. The technical knowledge necessary for use can be minimised through a well designed user interface. As they can be used constantly by a continual cycle of users, the much higher initial investment costs are warranted by this higher level of continued usage. The user can also potentially access either the Northern or Southern hemisphere, rather than the one where they are physically located, opening up more objects in the night sky to view.

Typically using these telescopes involves the use of either a request going into an online automated scheduler for a robotic telescope to return an image within a reasonable time (autonomous control, see Figure 3L), or through direct remote control of a telescope, located in a favourable time-zone (remote control, see Figure 3R).

\begin{figure}
\includegraphics[width=14cm]{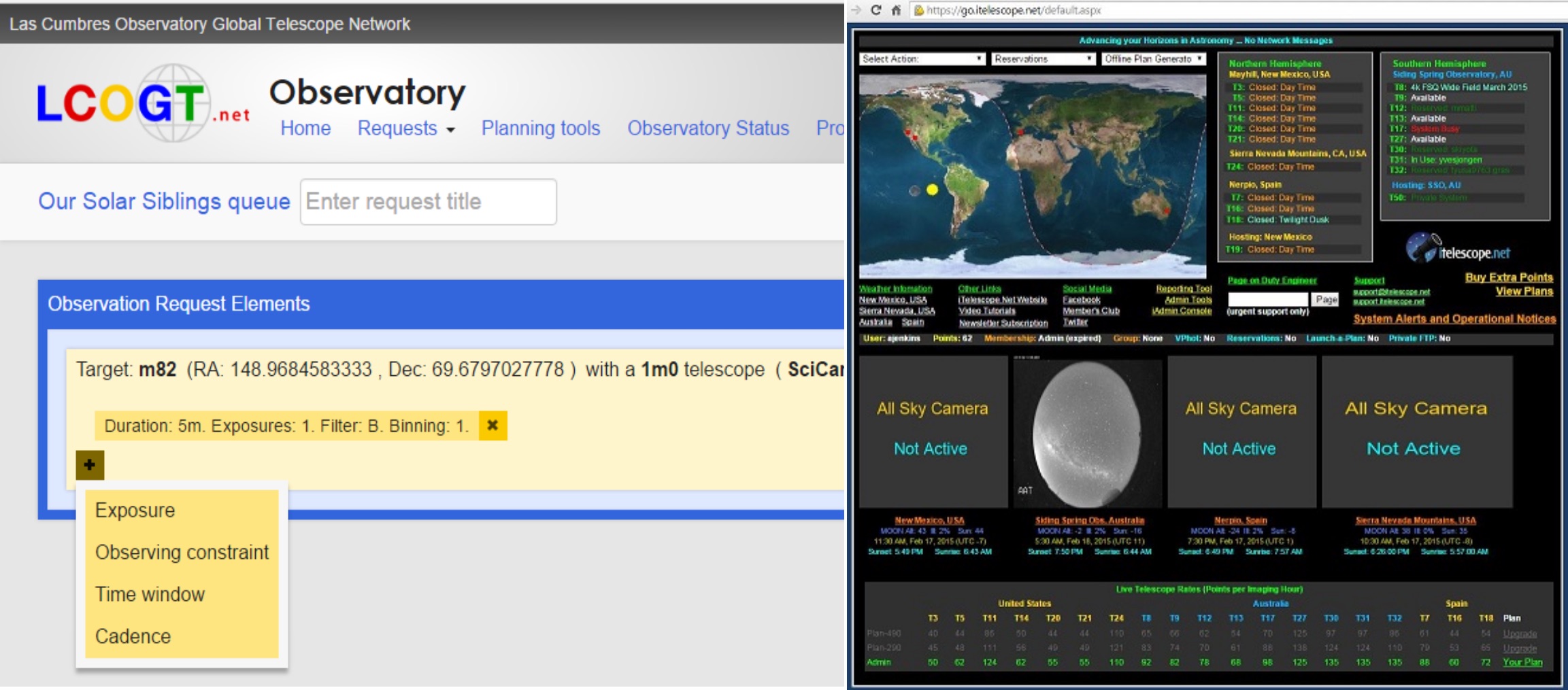}
\caption{Examples of robotic telescope user interfaces. On the left (L) is the Las Cumbres Observatory robotic scheduler interface. On the right (R) is the itelescope remote interface.}
\label{sample-figure}
\end{figure}

Robotic telescopes support new types of research that overcome human limitations. They can be set to monitor objects autonomously at every dark moment in remote locations repetitively and relatively cheaply over long periods of time, be driven by algorithms to maximise the efficiency of multiple observing programs as well as respond rapidly to urgent requests \cite{Eaton2003}.

Remote control observing is quite inefficient, solid estimates are rare, but Baruch \cite{Baruchetal2008} estimates that a robotic autonomous operator can work between 4 and 20 times faster than a human operator. A further step to make robotic observations more efficient, undertaken by the MicroObservatory telescopes \cite{Gouldetal2006}, is to only take a single image of a popular observation and send this one image out to multiple users, a method which is also employed by the Liverpool Telescope in their National Schools' Observatory program \cite{Steele2004}. Without sufficient context this could be perceived by the student audience as being something akin to a simple Google image search and not the product of their observation request.

\section{A Brief History of Robotic Telescopes in Education}

The earliest example of an automated telescope that ran over multiple nights was an 8" reflector telescope at the University of Wisconsin during the mid 1960s \cite{Code1992}. There were numerous other attempts prior to 1990 but not many of these achieved a stable functioning state. Notable exceptions to these were: the MIT Automated Astrophysical Observatory \cite{McCordetal1972}, Colgate's 30" \cite{Colgateetal1975}, the Phoenix-10 at Fairborn Observatory in 1983 built by Boyd, Genet and Hall \cite{BoydGenetHall1984}, the Carlsberg Meridian Telescope which began operation in 1984 \cite{HelmerMorrison1985}, the 0.8m Berkeley Automatic Imaging Telescopes \cite{Fillipenko1992} and the 30" Leuschner Observatory Telescopes \cite{Perlmutter1992}. These telescopes were largely focused on scientific research.

The Remote Access Astronomy Project \cite{Lubin1992a, Lubin1992b}, constructed as an undergraduate project between 1989 and 1992, used a 14" telescope with digital camera, on the roof of the UC Santa Barbara physics department which could be remotely driven. Teachers and students could access an image database, communicate and send observing requests to the telescopes by connecting to a bulletin board system via modem. Image Processing was undertaken through the custom-built software, IMAGINE-32 \cite{Lubin1992c}

The first research-grade instrument to be offered for educational use was the 24" telescopes at Mt Wilson, CA through the Telescopes in Education project \cite{Clark1998}. This project began in 1993 with students being able to speak to an operator while controlling the telescope. Smith \cite{Smithetal2001} describes a London high school's experience with TIE. Early experiences pre-internet involved direct dialing twice (once for voice, once for data) to the telescope which led to substantial cost in telephone calling charges. The tangible nature of hearing the telescope tracking added to the experience. The image itself would take 5 to 10 minutes to download.

During a similar time period, the Bradford Robotic Telescope \cite{Baruch2000}, as shown in Figure 4, was developed. The telescope control computers were attached to the internet via phone line and were sent a crude list of targets for the night, which one of the computers fine-tuned. While the instrument was initially built as an engineering experiment, the demand from the education sector was quite significant and surprising. The original telescope ran up until 1998 when a lightning strike decimated the electronics of the telescope. The telescope was insured and eventually replaced, the replacement runs until this day as part of an astronomical facility on Mount Teide, Tenerife with operations taken over by The Open University in 2016.  It provides time for free for education uses and runs a low cost service for amateur astronomers.

\begin{figure}
\includegraphics[width=14cm]{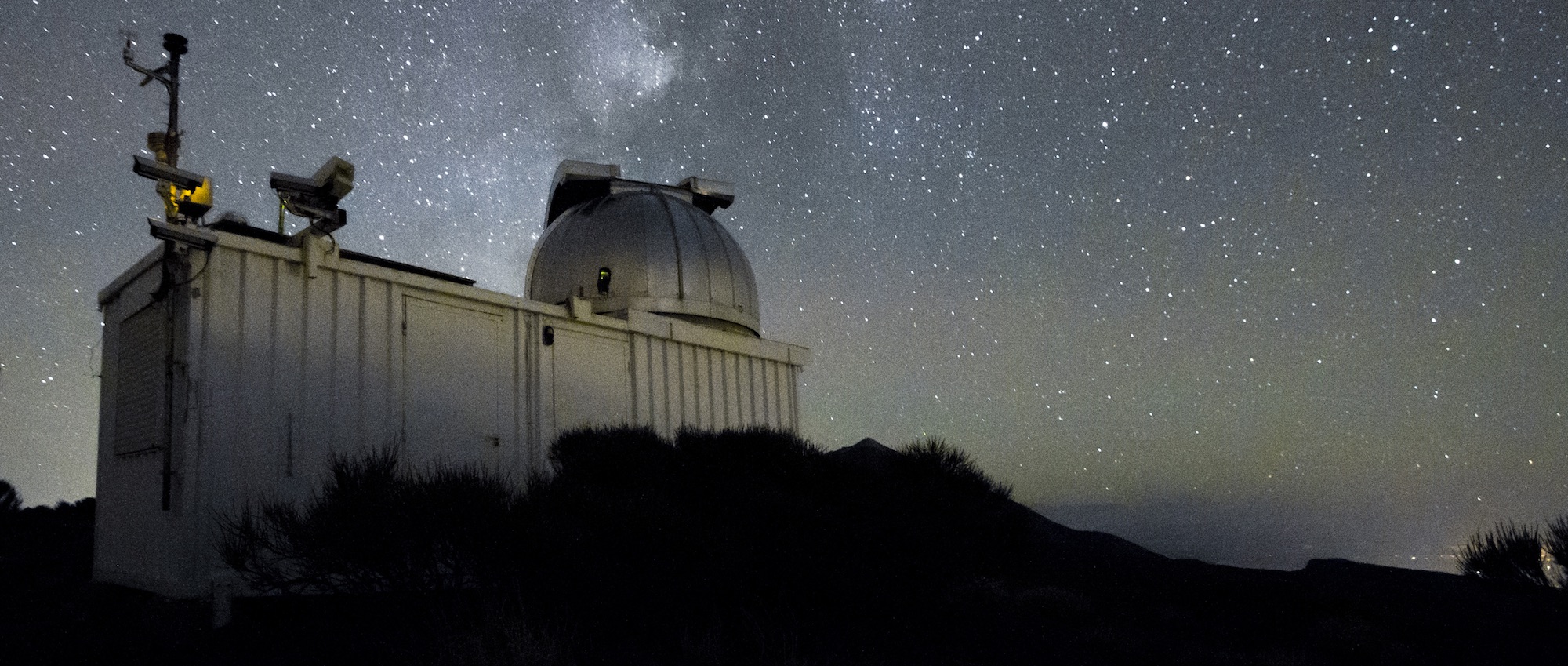}
\caption{An image of the Bradford Robotic Telescope, now the Autonomous Robotic Telescope (ART), at its site on Teide, Tenerife.}
\label{sample-figure}
\end{figure}

Hands-On Universe \cite{Pennypacker1998} was an early project beginning in the 1990s which did not necessarily center around a particular telescope but rather tried to create a network of telescopes for participating teachers and students to use, as well as providing teacher training, materials and software \cite{Pennypacker2003}.  The Yerkes 24" was roboticised and used for HoU, as were the 30" automated telescope at Leuschner Observatory \cite{AsbellClarkeetal1996}, the (then) 17.5" NF/Observatory in New Mexico (\cite{Neely1995}) and a variety of telescopes in Japan, France and the USA (\cite{Boer2001}). Currently HoU functions underneath the larger umbrella of Global Hands-On Universe (GHoU). A related collaborative project of institutions based in Europe (EU-HOU, \cite{Ferlet2008}) also operates as part of the Global-HoU. Early projects involving deeper scientific investigations included Asteroid Search \cite{Spuck1998} and Supernovae research \cite{Richmondetal1996} with more recent projects looking at transits and binaries \cite{Gould2010}.

MicroObservatory is a set of five 6" telescopes (see Figure 5L) that are located at Harvard College Observatory and the Whipple Observatory in New Mexico \cite{Sadleretal2001}. People can submit targets from a pre-defined list to be observed and are notified via email when the telescope has observed the object and downloaded the image. The MicroObservatory telescopes are still available for use today through ``Observing with NASA".

The Charles Sturt University Remote Telescope Project \cite{McKinnonMainwaring2000} began in 1999 in Bathurst, Australia. It originally focused on providing remote access to a 12" telescope (see Figure 5R) to primary school students in other timezones. This was quickly expanded to include middle and high school students \cite{DanaiaPhd} through the usage of the Journey Through Space and Time materials \cite{McKinnonGeissinger2002, Townsend2016}. Direct control of the telescope was achieved by students through the use of a Remote Desktop Protocol with an in-person presenter available for mentoring at all times. This project continues today as the OSS Remote Telescope Project utilising the same telescope as well as the SkyTitan telescopes in Wyoming and Texas.

\begin{figure}
\includegraphics[width=14cm]{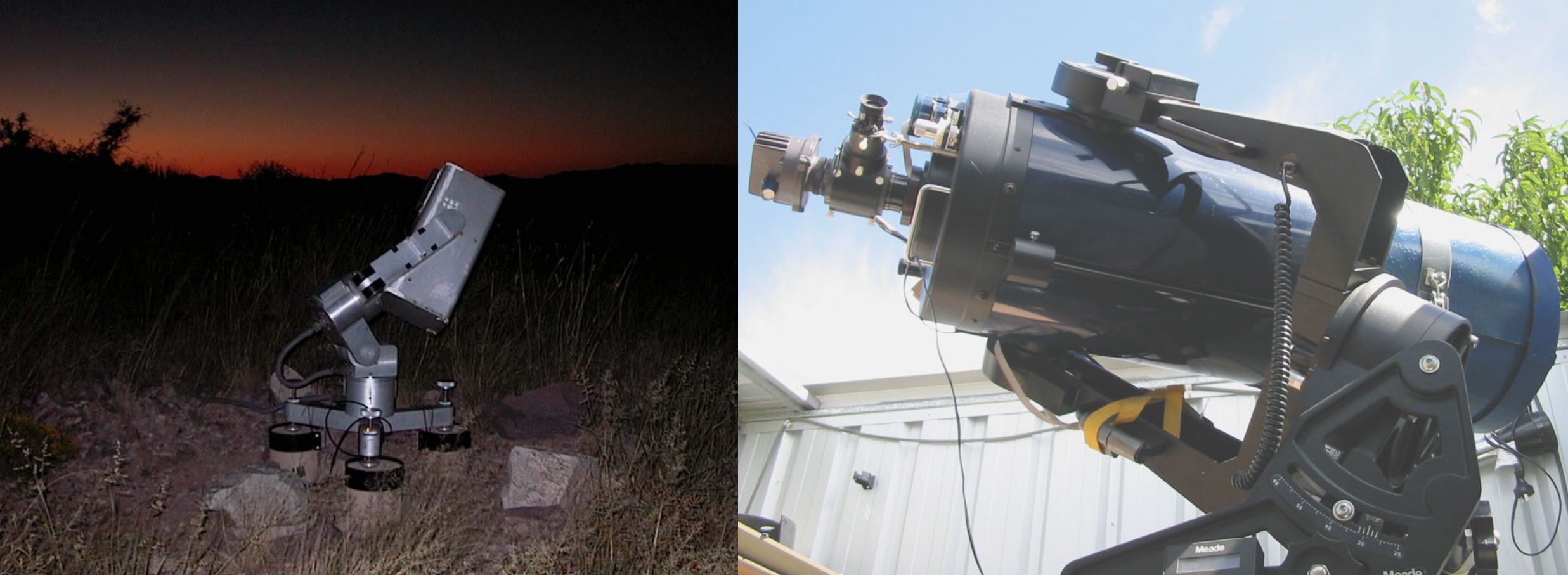}
\caption{On the left (L) is one of the 6" MicroObservatory telescopes. On the right (R) is the 12" Charles Sturt University Remote Telescope Project.}
\label{sample-figure}
\end{figure}

The Global Network of Astronomical Telescopes (GNAT) was an early attempt at a larger distributed telescope network beginning in the early-mid 1990s \cite{Crawford1996}. Telescope time and access was provided by membership and education was a prominent part of its mission \cite{Crawford1997}. The network was still hard at work in the R\&D phase in the mid-2000s \cite{Crawfordetal2003} and it seems to have been little active since 2009.

\section{Current Providers of Telescope time}

There have been a large variety of other attempts at small to medium size telescope networks. Quite a significant fraction of these networks do not reach stability over 3+ year timescales. Many of the endeavours mentioned in a 2010 survey \cite{GravesMackie2010STAR} of internet accessible telescopes are now, just over five years later, defunct or with uncertain status. Hence, our list is limited to those larger institutions who have shown relative stability, are currently functioning, and provide time explicitly for education. We will also not discuss the education programmes of observatories which have education and outreach programmes which do not involve capturing data from their telescopes (e.g. Herschel Space Observatory, European Southern Observatory).  Our list is intended to demonstrate the variety of programs and resources which are available, and not to provide a complete list of all available programs (see \cite{FitzgeraldHollowEtAl2014} for a more complete list). Inevitably our sample also has a bias predominantly towards English-language projects, as there are practical limits to how far investigate language specific programs, which we acknowledge.

The Liverpool Telescope \cite{Steele2004} is a 2 metre class telescope owned and operated by Liverpool John Moores University on La Palma. 5\% of the time available on this telescope goes to UK schools through the National Schools Observatory \cite{NewsamCarter2003}. The Liverpool Telescope has two sister telescopes, the  2 metre class telescope Faulkes Telescopes situated in Siding Spring Observatory in Australia and on Haleakala, Hawaii. All three of these telescopes were designed and built by Telescope Technologies Ltd. (now a subsidiary of Las Cumbres Observatory).

\begin{figure}
\includegraphics[width=14cm]{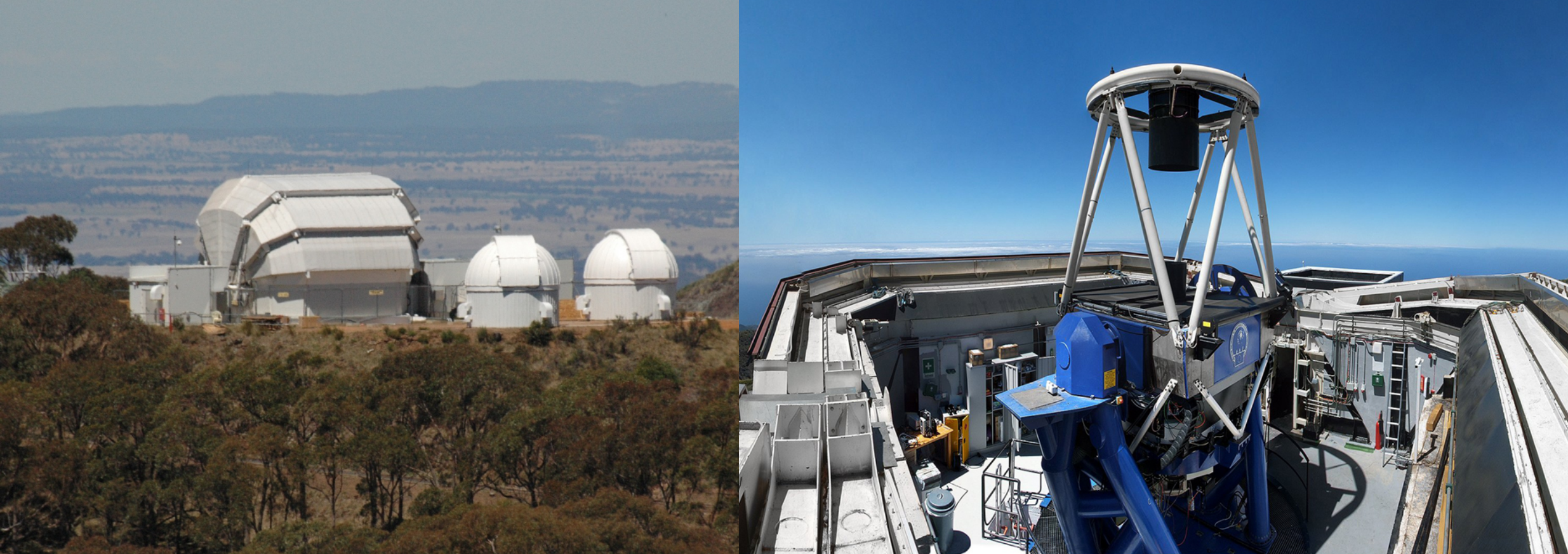}
\caption{The 2-metre class telescopes. On the Left (L) is the 2-m Faulkes Telescope South clamshell dome with two 1-m domes, owned by Las Cumbres Observatory. On the right (R) is the 2-m Liverpool Telescope clamshell opened to show the telescope itself. The Liverpool and Faulkes telescopes are very similar in original construction and design. }
\label{sample-figure}
\end{figure}

Las Cumbres Observatory \cite{Brownetal2013} is a global, homogeneous network currently consisting of the twin 2-metre Faulkes telescopes, nine 1-metre telescopes and seven 0.4 metre telescopes situated at various locations (see Figure 7) around the world (McDonald Observatory, Siding Spring, Cerro Tololo, Tenerife, South African Astronomical Observatory, and soon expanding to Ali Observatory in Tibet and Wise Observatory in Israel). Educational access to LCO is through a collaboration with different education partners around the world, each of whom manage their own education programmes using the LCO network. As well as being a service provider, LCO support a small number of informal education projects mirroring their research interests in time-domain astronomy.

\begin{figure}
\includegraphics[width=14cm]{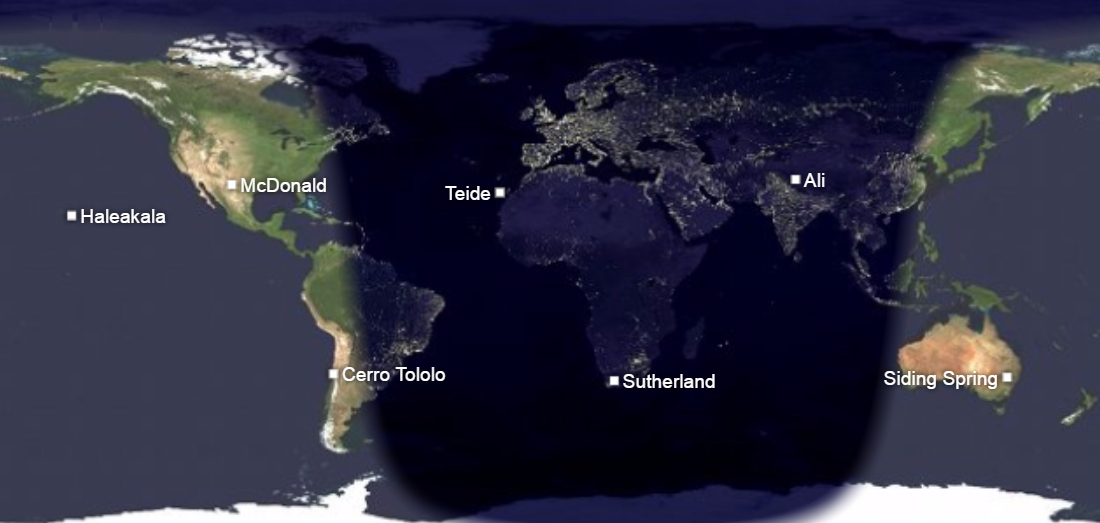}
\caption{Current observatory locations for the 18 telescopes in the Las Cumbres Observatory global telescope network.}
\label{sample-figure}
\end{figure}

SkyNet is a network of telescopes initially setup to observe gamma-ray bursts as six 16" PROMPT telescopes in Chile (see Figure 8) in 2004 \cite{Reichartetal2005} with four 17" PROMPT telescopes setup in Siding Spring Observatory in 2013 as well as a 24-inch ``Morehead Telescope" based on their home campus. They also provide access to a 20 meter radio dish and a variety of telescopes owned by other institutions, such as the 24" Yerkes or the GORT telescope at Sonoma State University \cite{McLinTFA}. SkyNet is also utilised in the Global Telescope Network (GTN) project \cite{McLin2014}

\begin{figure}
\includegraphics[width=14cm]{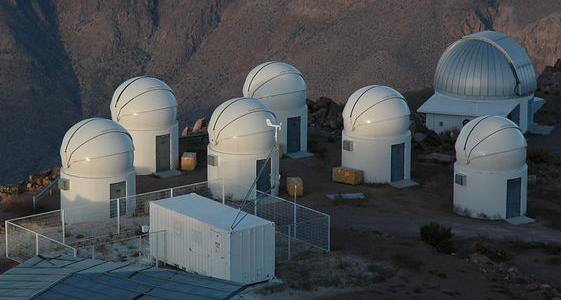}
\caption{The Chilean site of SkyNet showing the six 16" PROMPT telescopes.}
\label{sample-figure}
\end{figure}

iTelescope.net (see Figure 9) provides access to 19 robotic telescopes ranging from fairly small apertures up to 1 metre apertures in New Mexico, Spain, Australia and California for a paying membership. This value changes with the quality of the night (e.g. a lower rate applies when the moon is present).

The Sierra Stars Observatory Network \cite{Williams2010STAR} is a network of three roughly half-meter telescopes located in California, Arizona and New South Wales. They are designed to provide telescope time for education and research projects for a fee per hour. Further telescopes are intended to be added to the network over time \cite{WilliamsBeshoreTFA2011}

\begin{figure}
\includegraphics[width=14cm]{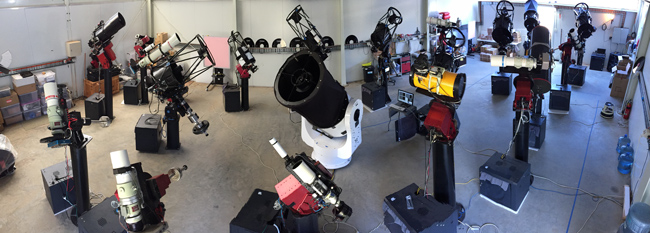}
\caption{The wide range of small to medium aperture instruments available through iTelescope. This is from their Siding Spring Observatory site.}
\label{sample-figure}
\end{figure}

SLOOH is a commercial network of ground-based telescopes in three locations, Teide, Chile and Australia. It was the first system that allowed users to observe night-sky objects in near real-time as a public outreach facility as well as providing more traditional robotic telescope access \cite{Whiteetal2008}.

The American Association of Variable Star Observers (AAVSO) runs a network called aavsonet \cite{Hendentfa2012} of ACP scheduled telescopes that accept proposals from their membership. It contains the two telescopes involved in APASS (\cite{Hendenetal2012AAS}) as well as four roughly half-meter telescopes and five small aperture telescopes for bright star use. These are located at various locations in SW USA, New Zealand and Chile.

MONET is a pair of 1.2m telescopes operated by the Georg-August-Univesitat Gottingen \cite{Bischoffetal2008}. One is located at McDonald Observatory and one at the South African Astronomical Observatory. While the telescopes have been around for a decade, they have been plagued by continuing technical issues but have had some success in providing time to student researchers.

Western Australia has two publicly accessible telescopes, the 1 metre Zadko Telescope \cite{Cowardetal2011} which aims to provide time for student asteroid and supernovae searches and a pair of telescopes (35cm and 43cm), called the 'SPIRIT' telescopes hosted through ICRAR \cite{Luckas2013} for general student use. Also in WA, there is a node of the Falcon Telescope Network, a currently growing network of 50-cm telescopes over 12 sites created by the US Airforce to track satellite and space debris positions. Some of the time available on the network will be provided to education users.

While most robotic telescopes deal with optical observations, there are also robotic radio telescopes available for use by education users. While radio observations lack the obvious aesthetic appeal of a crisp optical image, there are opportunities for students to undertake valuable scientific investigation using these telescopes. The Pulse@Parkes \cite{Hobbsetal2009} project uses the Parkes Radio Telescope in Parkes, NSW, Australia to directly observe pulsars with various school groups over the course of days and years. A similar project, without the direct control element, is the Pulsar Search Collaboratory \cite{Rosenetal2010}.

Small single-dish radio telescopes can be utilised for a variety of astronomical studies, such as brown dwarf variability, masers, gamma ray bursts, surveys and pulsars as well as a number of educational purposes \cite{Castelaz2003} The typical educational radio telescopes, such as the MIT Haystack design \cite{Salah2003}, consists of an antenna a few metres in diameter tuned to the 21-cm line. The Haystack 37-m telescope was also available for education users via remote control at 22-Ghz and 43-Ghz.

The Pisgah Astronomical Research Institute (PARI) has a 4.6m radio telescope (called 'Smiley') which offers remote usage through its School of Galactic Radio Astronomy. Skynet has a radio 20m dish at the National Radio Astronomy Observatory (NRAO) Green Bank site. The OpenScience Laboratory has a Radio Telescope called ARROW: ``A Robotic Radio-telescope Over the Web" available to use 21cm observations to map the spiral structure of the Milky Way.  European Hands-on Universe (EU-HoU) also has a variety of antennas intended to be used in schools distributed across Europe.

\section{Current Education Projects}

There are a diversity of programmes involving robotic telescopes, run by and with educational establishments across the globe. These programmes engage audiences from primary and pre-school level, through to university level courses and life-long learning. There are also a variety of uses in citizen science and public outreach. There are two broad uses of robotic telescopes for education in the various contexts, either engagement through simple astronomical imagery or providing access to collect research grade data, either astrometry, photometry or spectroscopy.

Colour Imaging is often used as a gateway into astronomy education using robotic telescopes. It has sufficient visual beauty to capture the attention of non-scientists \cite{Bessell2000} while still using authentic tools, such as FITS Liberator and image processing packages such as GIMP or Photoshop \cite{LindbergChristensen2007}, to work with relatively raw data. Very powerful visualization tools and techniques can be used to present and interpret astronomical imagery beyond a simple colour image \cite{Rectoretal2007} which can lead to a deep, engaging, emotional response on the part of the viewer \cite{Arcandetal2013}.

While this is likely the case, and colour imaging is the core of many optical robotic telescope education projects, there has not been a great deal of research on the efficacy, in terms of knowledge gain or engagement, on the student. It is a commonly held belief that colour imaging has merit amongst most projects and groups and, while unlikely to be shown dramatically false, this assertion still needs to be validated.

Other activities generally include more scientific uses of the instrumentation, whether astrometric, photometric or spectroscopic. The variety of these approaches is nearly as varied as the number of approaches in typical mainstream astronomy, albeit with a generally more limited timescale and scope. The following sections outline approaches taken in the three broad domains in which robotic telescopes are used, Formal School Education, Undergraduate Education and Outreach \& Public Engagement.

\subsection{Formal School Education K-12}

There are many reasons for including astronomy in the school curriculum, many of which were outlined by Percy \cite{Percy2009}. Some graduates will not have studied science beyond the upper high school level and this is typically the last moment the entire student body can be engaged, although the exact educational level varies from nation to nation. One of the major difficulties in providing experience to students at high school level is trying to fit an intentionally beneficial experience into the actual curriculum and syllabus. This becomes even more problematic when there is a mismatch between the intended spirit and wording of the curriculum, and the final curriculum documentation \cite{SlaterSlater2015}.

Interaction with telescopes particularly, helps to provide a different interpretation of the ``scientific method" involving observation, simulation and theory, than the typical  experimental textbook approach commonly provided, and can provide a basis to improve an otherwise uninspiring school experience \cite{Danaiaetal2013}. Robotic telescopes also provide strong links with other sciences, particularly physics, utilising concepts to do with light, gravity and instrumentation. Providing students with access to a means of looking at the size and age of objects in the Universe provides a less abstract/theoretical basis for studying time, distance and size scales. The general understanding of astronomy when students begin higher level (e.g. high-school, undergraduate) courses is generally fairly low \cite{DanaiaMckinnon2008, Slater2015b}. Students, at any level, typically also have very minimal knowledge of the night sky in terms of constellations and major stars \cite{Hintzetal2015}, and small practical knowledge of what cannot be seen with the naked eye.

The ``Build the Telescope and they will learn" approach \cite{Slater2014} is common throughout the literature involving formal school use of robotic telescopes. There is much literature on how such telescopes \textit{will} change science education but only a small amount (in comparison) has been reported about actual changes. It is not much of a leap to imagine that bringing such a new and alien technology into the everyday classroom is a significantly more complex and intimidating process than was originally imagined.

The last two decades have shown considerable success in terms of technical capacity but not in producing and supporting widespread inquiry-based astronomy education in formal education as was the usual intention \cite{FitzgeraldHollowEtAl2014}. It has been noted in various publications (e.g. \cite{Bretones2011, Slater2014, Fitzgeraldetal2015JEST}) that there has been a significant lack of deep research and understanding of how to translate this technology into the classroom.

\subsubsection{Some Examples of Formal School Projects}

The Bradford Robotic Telescope, currently in transition of ownership to the Open University, focuses on primary school levels providing access to its three cameras, which can be used to look at entire constellations (40$^{\circ}$), relatively large objects (4$^{\circ}$) and more deep sky objects (1/3$^{\circ}$). Teachers and their students autonomously control the BRT after a period of training, both through teacher workshops and in-school workshops. Then the teacher assumes the role of mentor to undertake astronomical investigations with their class. It was found that engaging primary school teachers in using the telescopes pushed the teachers to learn more about astronomy \cite{Baruch2011}. Is is claimed to have found a 30\% jump in applications in the STEM-based subjects from 50 schools that fed Bradford University compared to the 80 who did not \cite{Baruch2015}.

The original MicroObservatory telescopes \cite{Gouldetal2006} continue to be operational, providing images (cumulatively totalling in the millions) via autonomous control. This enables students to undertake many simple investigations on astronomical images, the most popular being the Moon. A new approach to using MicroObservatory is being trialled called ``Laboratory for the study of exoplanets" \cite{Gouldetal2012} (See Figure 10). This involves students observing known predictable planetary transits and undertaking photometry and modelling to estimate the size of the planet.

\begin{figure}
\includegraphics[width=14cm]{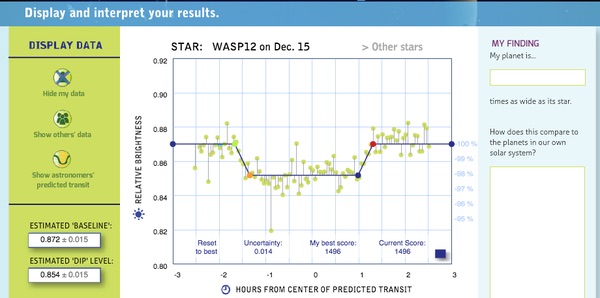}
\caption{A screenshot from the current MicroObservatory online exoplanet educational project.}
\label{sample-figure}
\end{figure}

The Faulkes Telescope Project (FTP) is a major education partner of Las Cumbres Observatory (LCO) network. It provides access, education and support to education users in the UK and beyond using the LCO network. It was initially started by Dr Martin Faulkes who provided \pounds10 million initial capital to create two 2-metre class telescopes available to schoolchildren in the UK under the slogan 'Real time, real science, real scientists' \cite{Beare2007}. FTP has continued as an official education partner of LCO, now that the Faulkes Telescope North and South are owned and operated as part of the LCO network. Previously they provided remote control of the 2-metres from the UK classrooms over the internet \cite{Roche2005}, although some participants could participate in more extended research projects via autonomous control \cite{Beare2004}. Now the telescopes are accessed primarily under autonomous control, through network wide autonomous scheduling, to provide access to high-quality imagery as well as data for a variety of research projects including variable stars, open clusters \cite{Lewisetal2010}, a variety of related education programs \cite{RocheFTPpdf} as well as recently, Gaia transient follow-up observations. Figure 11 shows examples of students and teacher participants in a supernovae-based school project.

\begin{figure}
\includegraphics[width=14cm]{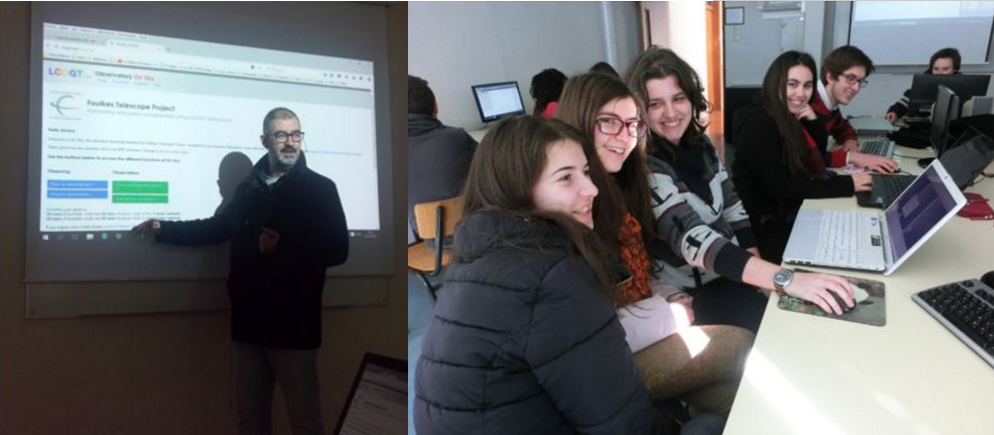}
\caption{High School students and their teacher interact with Las Cumbres Observatory telescopes and analyse supernovae data through the Faulkes Telescope Project.}
\label{sample-figure}
\end{figure}

The International Astronomical Search Collaboration (IASC) is an educational project that provides pre-observed data from a variety of robotic telescopes around the world to students in their classrooms (see Figure 12) to search for moving objects, primarily asteroids \cite{Milleretal2008}. Over 25 main belt asteroids have been discovered by students and thousands of measurements have been submitted to the Minor Planet Center.

\begin{figure}
\includegraphics[width=14cm]{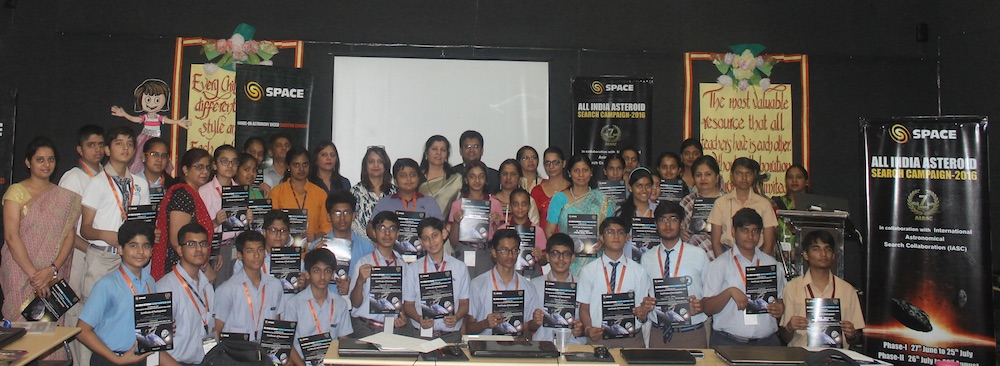}
\caption{An example of a class who participated in the IASC All-India Asteroid Search Campaign in 2016.}
\label{sample-figure}
\end{figure}

Our Solar Siblings (OSS) is a project based in Australia focused at bringing telescope access into the high school classroom to model authentic research. It focuses on the Year 10 Australian Curriculum, the final stage where science is a mandatory subject and where students are expected to learn about the components of the Universe and the Big Bang. Interested Students can then undertake their own authentic research-grade independent research projects, usually on RR Lyraes or Open Clusters (see Figure 13), mentored by project staff and professional astronomers. Early versions of the curriculum materials and general approach were also utilised by an earlier project primarily using the LCO 2-metre telescopes \cite{Danaiaetal2012}.

\begin{figure}
\includegraphics[width=14cm]{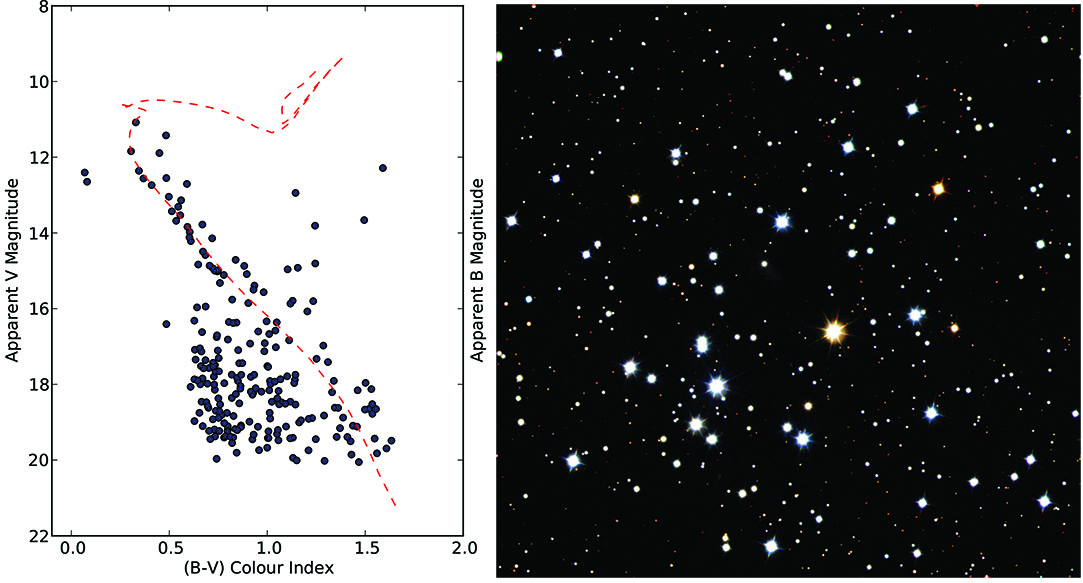}
\caption{An example of a student constructed image of an open cluster and its colour-magnitude diagram from the Our Solar Siblings project.}
\label{sample-figure}
\end{figure}

A variety of different projects in Hawaii' have utilised robotic telescopes for student research. From 1999-2004, the NSF funded a teacher enhancement workshop called ``Towards Other Planetary Systems" \cite{Kadookaetal2002} which reached 75 teachers over the five years, one third of them returning for future workshops.  Faulkes Telescope North, situated on Maui and operated by Las Cumbres Observatory beginning in 2005, continues to be used in a variety of contexts involving students and teachers (see Figure 14) with both amateur and professional astronomers \cite{Kadookaetal2008}. The HISTAR program, running since 2007, has provided students with astrometry and photometry skills to undertake authentic science projects using FTN, as well as a number of smaller robotic telescopes, to be presented at state fairs.

\begin{figure}
\includegraphics[width=14cm]{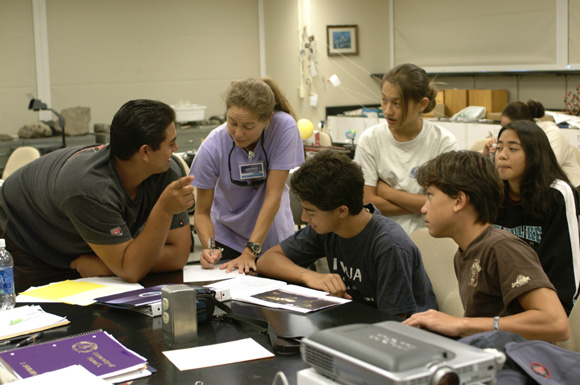}
\caption{An instructor with students participating in the HISTAR program in Hawaii}
\label{sample-figure}
\end{figure}

Project OBSERVE was a 5-year project (now completed) organized using the SkyNet telescopes with high schools (75 teachers, 5000 students) in the local area around the University of North Carolina. There are also programmes, such as Astronomy Live! run through UCLA or the long-running Summer Science Program, that take high school students on workshops using robotic telescopes over the summer break.

Las Cumbres Observatory, runs its various formal education programme internationally largely by providing time for official education partners to use, such as FTP and OSS above, as well as the Institute for Astronomy at the University of Hawaii, Universe Awareness from the Netherlands, Universe in the Classroom in Wales, McDonald Observatory in Texas and Astronomers without Borders. This model differs from others because it places LCO as a service provider (of telescope data) and the education partners as the content deliverers.

\subsection{Undergraduate Education}

In the United States, rough estimates are that 10\% of all undergraduate students will take an introductory astronomy course typically to meet a general education requirement in their undergraduate degrees \cite{PartridgeGreenstein2003}. In this particular country, this is a very significant proportion of the general population that would be touched by introductory astronomy, typically the last formal science these students will undertake \cite{Waller2011}. In other countries with differing tertiary education systems, this is typically not the case. For instance, in Australia, only a few thousand of the nation's 1.5 million undergraduate students will undertake an introductory astronomy subject \cite{Lazendic-Gallowayetal2016}.

While interest in using remote-control telescopes at the non-science major, high enrolment, ``astro101" subjects is large, where the telescopes would fit into such a subject is unclear and would require appropriate classroom-proven curriculum materials that are easy to implement \cite{SlaterASP2007}. Usage of robotic telescopes in such undergraduate courses is very uncommon. The problem of how robotic telescopes with finite telescope time provision could be used in classes with enrolments in the hundreds or thousands, such as a typical astro101 course, has not, as yet, been solved. An exception to this is SkyNet, which through its host institution runs a variety of undergraduate programs using the SkyNet telescopes to explore undergraduate content at an astro101 level.

Nevertheless, providing robotic telescope usage for undergraduates in non-astro101 contexts can provide a rich authentic experience. It can encourage creativity and discovery just at that point of their education career where the content can start to become quite dry and overwhelming \cite{Fillipenko1992}. For those students who continue their astronomy careers throughout undergraduate, the robotic telescope provides an important exposure to telescopes, providing insight to observational data that may be taken at face value otherwise \cite{Privon2009} and also potentially to increase their interest in a postgraduate science career \cite{Russell2007}.

In contrast to the high school projects and the major telescope institutions that serve them, undergraduate institutions have a much greater resource pool. This can sometimes lead to the existence of their own, more or less stable, remotely controlled observatory that they can use in their teaching programs \cite{Caton2003}. These smaller facilities also allow exploration of more speculative or lower expected impact research programs (e.g. \cite{Im2015, Dukes2003, Eastwood2003}). While their, generally institution-bound, location likely significantly decreases the quality of the observatory site, it does provide capacity for students to directly see the engineering and instrumentation side of telescopes.

Those students who may be continuing on towards higher level courses or even a scientific career may want to dig a little deeper into astronomical research during their undergraduate experience. Amongst the most notable and long-term series of undergraduate student research programs is Percy's work through the University of Toronto Research Opportunity programs over many decades outlined in Percy \cite{Percy2008b} with many examples of student publications over the years.

Beaky \cite{Beaky2010STAR} outlines a few broad principles for undergraduate astronomy research. These include that 1) the research must be sustainable, in the sense that accumulated experience is passed from one student to another before graduation, 2) that the technology must be reliable and relatively easy to use, 3) that the research must be designed to lead to publishable outcomes if it is to be truly called 'research' even if it does not succeed in publication and 4) that the undergraduate research needs to be integrated into the curriculum.

There are many examples of programs within which single university-driven robotic telescopes can, and have, been utilised in an undergraduate context. A good example of this is the ``Physics Innovations Robotic Astronomical Telescope Explorer" or PIRATE \cite{Holmesetal2011} which is a 17" telescope (see Figure 15) run by the Open University. The original intention was to provide a hands-on experience to groups of 2-4 distance education students who collaborate via the internet remotely through the institution with limited remote supervision \cite{Kolb2014, Brodeuretal2014}.  Targets selected by third-year observing groups largely involve periodic variables \cite{Nortonetal2007}, sometimes leading to scientific papers in the pro-am journal, JBAA \cite{Pulleyetal2013, Bruceetal2013}. Second year students generally create colour magnitude diagrams of open clusters.

\begin{figure}
\includegraphics[width=14cm]{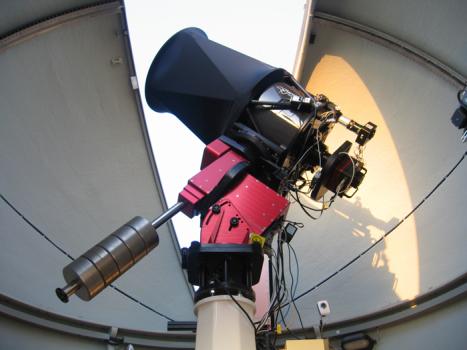}
\caption{The 17" PIRATE Telescope run through The Open University, while located in Mallorca before moving to Tenerife in 2016.}
\label{sample-figure}
\end{figure}

Another example is the work done through a variety of institutions including the Boyce Research Initiatives \& Education Foundation (BRIEF), Concordia University in Irvine, Cuesta Community College, California Polytechnic University in San Luis Obispo and the Institute for Student Astronomical Research \cite{Genetetal2010}. These are one semester astronomical seminars at community colleges and undergraduate which also include high school students looking for future course credit. The topic is generally double star observations \cite{Genetetal2012} but has also included cepheids, asteroids and exoplanets. The goal of the seminar is for students to create a peer reviewed published article for the amateur/pro-am literature.

Other examples include the Keck Northeast Astronomy Consortium \cite{Partridge2003} which has been running for nearly three decades providing astronomical research to eight small liberal arts colleges and universities in NE USA.  The 0.51m Iowa Robotic Observatory at Winer Observatory utilises their telescope in introductory, major-level and upper-level courses. Asteroid Lightcurves \cite{Willis2004}, a search for extra-solar planets around white dwarves and a variety of other projects are undertaken. The 20" Northwest Indiana Robotic (NIRo) Telescope provides time for undergraduate research and education at Purdue University, Calumet. Monitoring of exoplanets, blue stragglers in open clusters, quasar variability and asteroids are some topics that are explored.

There are other organisations that are not specifically astronomy which also provide some scope for student astronomical research with robotic telescopes. Two examples are Research Experiences for Undergraduates which provides research experience with projects run through the National Science Foundation or the Council of Undergraduate Research which has a number of projects and approaches to promote undergraduate student research. These two are examples from the United States, but there is likely similar national programs in many other countries.

\subsection{Outreach and Public Engagement}

While formal astronomy education takes place within classrooms, there are many other places, such as planetaria, museums, astronomy groups, newspapers, television, radio, the internet, books, magazines, youth groups on hikes \cite{Fraknoi1996} where informal astronomy education can take place. It is part of the ``astronomers burden" \cite{Caton2003} that astronomy is highly visible, compared to subjects like chemistry or pure mathematics, and of inherent interest to the public. The likelihood of students being interested in pursuing astronomy beyond high school is significantly correlated to their participation in these extracurricular astronomy or space science activities, such as observing stars or reading/watching science fiction \cite{Bergstrometal2016}. All outreach and engagement projects can have a lasting impact if they are also accompanied by educational resources and a dissemination plan.

By comparing the situation to that of a successful school sports league, Gelderman \cite{GeldermanPasaIAE2008} argues we need to provide out-of-school time astronomy students 1) open and easy entry level recreational opportunities, 2) a series of challenging activities with steadily increasing challenge and reward, 3) high-level competitive experiences in an astronomy club to help drive further participation in extracurricular astronomy. This can model similar approaches to those listed in the Formal Education or Undergraduate sections but to greater depth. SkyNet Junior Scholars, a good example of this approach, is run primarily for non-formal education such as youth clubs, after school programs, museums and camps. Professional learning about using the SkyNet telescopes is required for these group leaders who can guide youth explorations using the telescopes.

The field of public outreach and  engagement using robotic telescopes employing a more direct interaction between robotic telescopes and the general public is less well developed. There have been some attempts at providing museum portal access to optical telescopes. The Hands-On Universe project gained funding to attempt a museum exhibit to allow visitors to control two remote telescopes \cite{Pennypacker2004}. SkyNet has also provided in the past a web-based museum portal where visitors can request an image that is taken later by the network and sent to their email address. There are many undocumented examples of robotic telescope groups introducing informal audiences at exhibitions and festivals, and the perception by the public is highly dependant on the individuals performing the demonstrations.

The telescopes provided by SLOOH and GLORIA \cite{Whiteetal2008} have been used to live stream astronomical events, notably the 2012 Solar Eclipse, viewed through one of their remote telescopes to a wider non-specialist audience.  These provide excellent opportunities for outreach as the phenomena themselves gain significant media attention in their own right.

During International Year of Astronomy 2009, European Southern Observatory hosted an ambitious event, "Around the World in 80 Telescopes" \cite{PiercePrice2009}. During a continuous, 24 hour webcast, hosted by a team of 6 presenters in ESO headquarters in Garching, Germany, the audience were shown live video feeds from telescopes from across the globe. While all of the telescopes involved were robotic, the majority of them were remote controlled from the same site, with the video feeds coming from the control rooms. This event attracted an audience of 156,000 viewers over the 24 hour period, with a maximum of 3600 simultaneous viewers. Although this audience did not play an active part in the outreach, it still had a large reach and hopefully a significant impact.

Twitter has been used by many different groups and individuals for astronomy outreach, virtually from its beginning. The idea of telescopes (and other scientific robots, such as Mars Curiosity rover and Planck Satellite) broadcasting their positions, targets and even developing 'personalities' was hugely appealing to the general public \cite{Lowe2009}. Some even chose Twitter above more traditional media outlets to broadcast their status and discoveries, such as ESA Rosetta's ``Hello World" Tweet, announcing it had successfully awoken from hibernation. The idea of using Twitter and robotic telescopes for outreach was taken further by Las Cumbres Observatory, where in 2011 they partnered with Irish comedian Dara O Briain, who live tweeted images and commentary from an observing session on one of their 2-meter telescope, to his 2 million followers \cite{Gomez2014}, in an event called \#ShowMeStars.

LCO Asteroid Tracker is a recent attempt to get the public to trigger their own observations of an asteroid which is be added to a dataset for analysis and a later video for exploration by the public with an attached story about the particular object. For the non-astronomically inclined this type of activity can be a meaningful contribution with an engaging story and requiring minimal technical knowledge.

\section{Authentic Astronomical Research produced by Education and Citizen Science Users}

While the cutting edge of extragalactic astronomy at large redshifts is dominated by the extremely large (8m+) telescopes, there is much research that can be undertaken with smaller telescopes. There is some debate to the relative worth of small and large aperture telescopes with different studies coming to different conclusions from different metrics \cite{Sage2003} such as number of publications and citations \cite{Abt2003}. There is some research that is more suited to small-size (0.5m to 1m) and middle-sized telescopes (1m-3m), in particular topics such as stellar variability, Gamma Ray Bursts, Near-Earth Objects, Supernovae and exoplanets \cite{QuerciQuerci2000}.

It is also noted that many recent major scientific studies, such as 2MASS or SDSS, have been undertaken on modest sized telescopes \cite{Huchra2003}. The contribution non-professionals can make to astronomical research evolves over time as technology and techniques change. As most astronomers move to investment in larger aperture, highly competitive, instrumentation at the expense of smaller instruments and, as telescopes of modest aperture become increasingly available at a lower cost, pro-am organizations and individuals can step in to contribute to the scientific endeavour \cite{FeinbergPASAIAEpdf}.

\subsection{Citizen Science}

In astronomy the term Citizen Science is often used synonymously with the multi-project Zooniverse \cite{Fortson2009}, typified by the original project in this suite, Galaxy Zoo \cite{Lintott2008}. Although the majority of Zooniverse projects use large, pre-existing datasets in the field of astronomy there have been a few which were responsive to new data acquired by robotic telescopes. The projects, Supernova Zoo and Asteroid Zoo received data at regular intervals from Palomar Transient Factory and Catalina Sky Survey (respectively). While providing useful classification statistics, the interesting aspect of these projects is that the robotic telescope data was being used for a dual purpose of training a computer learning algorithm. The ultimate aim of the project was not to achieve a specified number of classifications but to eventually train the algorithm to accurately detect transient objects, eventually rendering the human interaction unnecessary.

While Zooniverse projects tap into and feed the motivations of the public to contribute to science and inspired some to dig deeper or to begin postgraduate research \cite{Graham2015, Mankowski2011, Raddicketal2010}, it is for the most part citizens volunteering to be basic interpreters of images rather than being more participatory in the larger process. Their methodology is to provide a large data-set to an audience without specialist skills and use the human brain's innate ability to recognize patterns to produce a statistically significant classification of the data.

The authors of this paper note that the Zooniverse approach is not the definitive citizen science method. We broaden the definition saying that citizen science should be an activity through which the participants will grow their skills and knowledge of the science area by being involved in the citizen science project. This inevitably requires more education and thus blurs the line between what is considered education, citizen science, and authentic research. The education involved can be simple and built into the interface used for the analysis of the data.

Citizen Sky was a large project run by the AAVSO on epsilon Aurigae as a focussed project for the International Year of Astronomy \cite{Price2012}. The goal was to discover the cause of this binary's eclipse. At the end of the project it was theorised to be an F-type star creating a disk shrouding another comparable mass star or star/s in the system \cite{Stencel2011}. Typically most observations were in the V-band made by 26 observers around the world using a variety of tools, telescopes and techniques \cite{Hopkins2012}. These were largely with equipment owned by the observers, but is a good example of a potential collaborative project accessible to the public that could utilise the robotic telescopes coming on-board now. Projects, like IASC and Citizen Sky, that include active robotic telescope use require providing a different framework for the public to incorporate their original observations and deeper analysis.

\subsection{Authentic Research}

Authentic astronomical research utilising high grade instrumentation can engage gifted or motivated students who otherwise may not be enthused by their typical everyday science lessons to consider the option of a potential STEM career \cite{HollowPasaTLA}. Many of these student research projects can be undertaken in the context of national schemes or competitions, extra-curricular astronomy and science clubs or external mentoring by a professional astronomer \cite{Hollow2000}. For those undertaking authentic research projects, students who undertake research in high school are more likely to be employed in a science-related career than those whose first research is in undergraduate \cite{Roberts2009}.

Particularly in the United States, there are science fairs at district, state and national scales which have also provided an avenue for communication of science by students, although the historical observational non-experimental nature of astronomy can be an impediment in this context \cite{Percy2003}. Variable Stars, in particular, have been a popular topic for student research projects due to the relative simplicity of the measurements compared to other more abstract fields of research \cite{Percy2006e}.

Many projects over the last two decades have claimed that authentic, original research on behalf of the students have been a core motivation in their projects but not as much progress or success as was hoped occurred on this front \cite{FitzgeraldHollowEtAl2014}. Part of this challenge is that the balance between authenticity, challenge and time within the curriculum is hard to achieve.

Despite this, some research that is undertaken within these projects can reach the level of publication in mainstream and pro-amateur scientific journals.  While not intending to be an exhaustive list, we present a sample of some of the students who have published in astronomy journals involving use of remote telescopes in an educational context. We recognise that this sample will likely be biased to those publications in larger projects from an primarily english speaking background.

Two students at Oil City High School in Pennsylvania participating in the Hands-On-Universe project were involved in the observations of SN 1994I in M51 \cite{Richmondetal1996, Pennypacker1996}. Students at Northfield Mount Hermon School, Massachusetts discovered the 72nd Kuiper Belt Object 1998FS144 \cite{Pack2000} with confirmation by students also from Oil City High School.

The MONET project produced a variety of papers with German Gymnasiums involving cataclysmic variables \cite{Beuermannetal2009, Beuermannetal2011} and  binary stars \cite{Backhausetal2012}. Students involved with the Space To Grow project \cite{Danaiaetal2012} produced papers on a planetary nebula \cite{Frewetal2011}, RR Lyraes \cite{Fitzgeraldetal2012} (see Figure 16R) and on a neglected open cluster \cite{Fitzgeraldetal2015AJ}. Through the NITARP project (see Figure 16L), students were involved with observations of the Witch Head Nebula \cite{Guieuetal2010}, cataclysmic variables \cite{Howelletal2006}, accretion discs \cite{Howelletal2008} and young stellar candidates \cite{Rebulletal2011}.

\begin{figure}
\includegraphics[width=14cm]{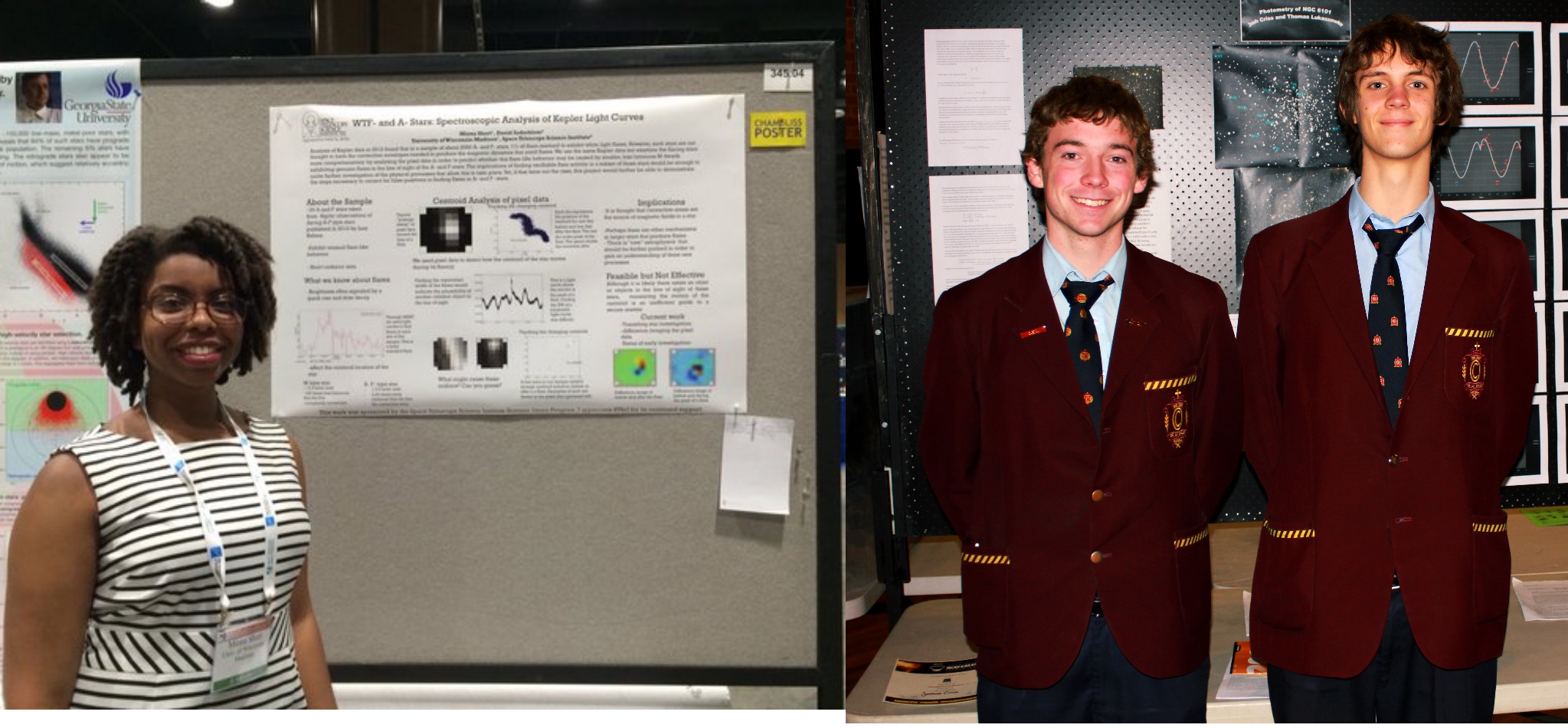}
\caption{On the left (L), a teacher involved in the NITARP project presents her research from a year-long programme. On the right (R), high school students present their results from their use of the LCO 2-metre telescopes in the Space to Grow programme.}
\label{sample-figure}
\end{figure}

Some student research has ended up in the amateur literature, recording scientifically useful measurements and discoveries that may not warrant a full publication in the mainstream professional astronomy literature. The journal of the American Association of Variable Star Observers (jAAVSO) has published numerous articles based on the work of students, both undergraduate and high school, on variable stars over the years (e.g., \cite{Percy2008a, Percy2008b}). Potential other places to consider for future publication of variable star related results include the  the professional Informational Bulletin on Variable Stars (IBVS).

Contributions of asteroid measurements to the Minor Planet Center by students are relatively common with many of these provided by the IASC project \cite{Milleretal2008}. The Minor Planet Bulletin contains much pro-am research surrounding asteroids, in particular asteroid rotation lightcurves. More generally the Society for the Astronomical Sciences (SAS) publishes a yearly proceedings that includes astronomical results. The Journal of the British Astronomical Association (JBAA) also publishes amateur/pro-am papers. The Research Based Science Education journal ran through the NOAO from 1999 to 2010 provided a place for scientific reviewed papers written by the high school teachers and students participating in that project. Results were from a variety of objects including Open Clusters, T Tauris, Variable Stars, Solar Observing, Galaxies and Nova.

A particularly accessible field to undergraduate students has been the field of Double Star Observations. A variety of publications has been published in the Journal of Double Star Observations \cite{Clark2010STAR} which can lead to inclusion in the Observations Catalogue of the Washington Double Star Catalogue \cite{Mason2001} or the Circular of Commission 26 (Double and Multiple stars) of the IAU. Monitoring these stars provides measurements for future astronomers to estimate the masses and absolute positions from their orbital ellipses \cite{Argyle2012}.

For Undergraduates, there are a variety of specifically undergraduate focused journals, such as Reinvention, Journal of Undergraduate Research and Scholarly Excellence, the American Journal of Undergraduate Research and quite a number of journals sharing the name ``The Journal of Undergraduate Research". The various levels of astronomy journals are probably still a better choice for astronomical undergraduate research publication. The Council on Undergraduate Research also provides a significant amount of support for organising and driving undergraduate research programs.

\section{Challenges and Opportunities}

\subsection{Quality of available supporting material for robotic telescope usage.}

Percy \cite{Percy2003} envisioned a robotic telescope farm facility, similar to the large networks available now, that had the potential to serve tens of thousands of students. He did stress that they need to be supported by a \textit{carefully-developed} set of curriculum materials and approaches. This need for quality materials has also been expressed for other astronomy areas such as the Virtual Observatory \cite{White2001} and astronomy education in general \cite{Russoetal2015}.

A typical student in an average classroom, while having seen spectacular images of the night sky perhaps on Google or on any form of media, tends not to have a great grasp of the types of objects out there in the universe. Placing a student in control of the telescope immediately without some support will not typically result in beneficial outcomes, they typically require some scaffolding \cite{Slater2014}.

There is a large amount of material of widely variable quality available from a variety of projects and repositories. There is certainly a fairly strong case that there is a lot of reinventing the wheel occurring with a lot of overlap between the material for all of the different robotic telescope projects. There are numerous classroom sets and training manuals on variable stars, open clusters, asteroids and colour imaging amongst other topics available online. In the future, perhaps collaborative efforts can be made to take the best of what has been learned into a more centralised location for all robotic telescopes, perhaps through online repositories like NASA Wavelength \cite{Cobbetal2015} or through online peer-reviewed repositories like astroEDU \cite{Russoetal2015}.

\subsection{Efficient use, reliability and scalability of telescope resources.}

There are only a finite number of robotic telescopes in the world and many more teachers and even more students. Assuming an 8 hour night, the absolute maximum a single telescope can provide is about 2100 hours assuming 75\% of the nights are available per year.  Assuming a total exposure time per colour image for the full set, of 10 minutes on a 1-metre telescope (for reasonable signal-noise ratio of an extended, deep sky source), and if the telescope was entirely dedicated to this usage, then $\approx$ 12,000 image sets could be collected per telescope per year.  Assuming roughly 750,000 students in a single grade of high school in the United States. This would require around 60, 1-metre class telescopes (or more of smaller aperture) fully dedicated to this usage so that each student could get one set of BVR images of their own at some point in their high school careers.

Just on these rough numbers, then about 1 percent of the necessary telescope time is available for such use. This is prior to including the vastly more lengthy exposure times for independent research projects to the tally. How are these limited robotic resources going to be distributed amongst this large set of users? Telescope time is a precious commodity, especially on research grade instrumentation.

Creative suggestions to overcome these time limitations early on included taking a single image per night, instead of taking a single image for every single request, and then distributing that single image to multiple users \cite{Sadleretal2001}. This may be a good solution where ownership of the data is not core to the program or for purely inspirational purposes, perhaps in a primary school or museum outreach context. It does however, take away a large motivation of providing access to students for authentic research purposes. While this potentially maximises telescope time, it takes away from the sense of ownership that may be the most important part of imaging with such telescopes.

There is some concern that teachers and students will not trust that the images are actually their own rather than simply just images that were taken at a prior date \cite{Gershun2014}, this is a problem that can be overcome with remote real-time control telescopes where users can test this hypothesis directly \cite{McKinnon2008} although the current availability of such approaches is relatively small but growing.

\subsection{Scalability of human resources}

What may be more surprising is that the limitations preventing larger groups of users to collect their own images from telescopes is far from the technical one mentioned above. From formal and informal discussions with many education and technical projects, actual recruitment and retention of education users of robotic telescopes is known to be a very tricky, very drawn-out, process. A recruitment of a teacher in one year does not transform into an in-class user until at least the next year due to school planning cycles.

Designing efficient user-friendly portals to access the telescopes as well as scalable and enticing training, and using professional learning methods is far more important at this stage. Scalable large networks such as SkyNet, Las Cumbres Observatory, iTelescope and others, are now in the domain where providing more telescope time is largely a case of providing more money to the technical institution. Scalable educational support, training and retention needs are currently trailing significantly behind the scalability of the technical instrumentation. This is equally important where teachers facilitate learning, as it is in self-directed, or IBSE, scenarios (where students direct their own learning).

For robotic telescope usage to really take off in the general teaching population, the robustness and reliability of the networks is the most important issue.  For example, if a classroom teacher requests an image in a certain time-frame so that it is ready to present to a scheduled class then, barring weather and mechanical fault, that image needs to arrive in the requested time-frame. If it does not, the teacher is unlikely to utilise the robotic telescope in class again. The technical provision really does have to be bulletproof \cite{Fitzgeraldetal2015JEST}.

\subsection{Instructor Barriers}

Even early on, it was noted that there were significant issues with being able to match user needs and their interests to the growing availability of this technology and that robotic telescopes were not necessarily a good tool for educational good, just because they were new and available \cite{Marschall1996}. Problems of user confidence in turning the classroom technology into an effective classroom tool saw some instructors struggling to provide it as an in-class tool, especially as it was not clear how to incorporate it. \cite{Gershun2014}

Even those instructors who are identified as positively inclined or 'alpha' instructors, while being aware of online resources, rarely made common use of online data for authentic inquiry in the classroom. \cite{Slater2009}. These results suggested that instructors required training that translates research into classroom ideas and well designed data-driven inquiry products. Schools and universities typically also have limited resources and facilities that are not low monetary cost will likely not have a large uptake and that uptake will only be by those who can afford it, thereby largely negating the 'equity of access' benefit, as far as socioeconomic indicators are concerned.

There are a number of key barriers for instructors in the uptake of robotic telescopes \cite{Fitzgerald2016Barriers} and inquiry-based teaching \cite{Marshall2016}. Primary school teachers largely have little to no astronomy background or even science at all and commonly hold just as many misconceptions as the students do \cite{DanaiaPhd}. High school teachers are equally likely to have little astronomy background  and, while generally required to have science as part of their undergraduate degrees in most countries, are usually biology or chemistry focussed with physics-related backgrounds being a much distant third (e.g. \cite{McKenzieetal2008}).

While it is easy to find those teachers who are already pre-disposed to astronomy, who probably already have a semi-amateur, amateur or even pro-amateur background, it is a very different story trying to recruit teachers for whom teaching astronomy and/or physics may be an uncomfortable period of time during their school year. This is a common problem with external science education programmes (in fact most innovations, e.g. \cite{Rogers2013}) where early success is achieved with the early adopter populations but recruitment stalls when attempting to reach the less positively inclined majority population of instructors.

To deal with this, many more instructor concerns must be identified and addressed with the majority of instructors, not just those who are already very keen. These include, but are not limited to, the lack of preparation time available, the lack of high quality and easily accessible professional development, capacity to fail without repercussions, unrealistic expectations on behalf of project personnel, lack of administrative or supervisory support, a lack of available funds, impenetrable and unreliable information technology systems and the common mismatch between what is expected by the curriculum and what is asked by project personnel \cite{Fitzgeraldetal2015JEST}.

Some approaches to dealing with instructors confidence is to provide authentic research experiences (e.g. \cite{Rebulletal2016AAS, DresnerWorley2006}) to increase their understandings of the nature of science. Prior experience with research has been shown to have the strongest correlation to accurate knowledge of the nature of science and scientific inquiry on behalf of the instructor \cite{Buxner2014}.

 \subsection{Image and Information Technology Issues for Educational Use.}

Images that come out of these robotic telescopes, if provided in authentic form, are generally Flexible Image Transport System (\cite{Wellsetal1981}), or ``FITS", images. This is not a familiar file format to teachers or students. Most of the software developed to undertake authentic science grade manipulation of these files are written for the Linux operating system. In contrast, essentially all education users will be using Windows or Mac OS based operating systems. These users also will lack the command-line/terminal and editing skills necessary to drive a lot of this manipulation.  SkyNet have attempted to overcome this issue by providing a web-access image reduction and analysis tool for their project called Afterglow run from a central server over the web. This is an approach also taken by Boyce Astro through the Boyce-Astro Research Computer (BARC)

Typically a raw image from the telescope needs to undergo a series of fairly standard image 'reduction' steps. The most fundamental of these are the bias, dark and flatfield subtraction. Thankfully, as these are heavily tied to the nature of the telescope and camera themselves, these steps are typically taken at the observatory before provision to the user.

While these basic reductions generally do take place, there are other image artifacts that need to be dealt with. Hot or cold pixels, column defects and cosmic rays need to be identified and removed. Unfortunately, there is no relatively user-friendly, accessible or free software that can remove these quickly and automatically. Fortunately, these issues can be very easily automated in Linux by the providers of the telescope access.

In recent years, placing the correct Right Ascension and Declination coordinates into the FITS file to accompany just the basic x,y pixel locations has become quite easy with the advent of astrometry.net software \cite{Langetal2010}. This is another step than can easily be undertaken at the provider's end.

For the most basic usages, such as examining the FITS file, stretching the image and making very simple measurements there are a variety of freely available software. FITS Liberator is the most useful piece of software for simple examination and advanced stretching in preparation for colour image processing. Subaru Makali'i Image Processor is a popular, very simple, very user-friendly Windows software that does basic stretching, blinking, photometry and astrometry.

SAOImage DS9 is a popular scientific viewer amongst professional astronomers which has native versions for Mac OS and Windows. The Windows version has not been updated since 2013. The original MicroObservatoryImage software is still available online, although it is a bit dated. SalsaJ \cite{Doranetal2012} was developed by and extensively used in the HoU and EU-HoU projects but as it is not under active development can suffer from software glitches as well as mildly inaccurate photometry results. It is based on the popular, cross-platform software, ImageJ \cite{schneider2012}.

To undertake scientific grade astrometric, spectroscopic or photometric measurements, analysis should be undertaken with a professional grade tool. For astrometry, the most popular of these is Astrometrica which is shareware and has an extensive trial period. BdW Publishing's Canopus is a mid-range option. For those who have monetary resources to expend, there are relatively high cost software packages such as DC-3 Pinpoint, TheSky/CCDSoft and Prism Pro that will undertake quality astrometric measurements.

For aperture photometry, the most robust freely available tool available is Aperture Photometry Tool. The interface can be daunting for first-time users, but it is heavily validated \cite{Laher2012a, Laher2012b}) and has successfully been used in research that has been published in the mainstream and pro-am literature \cite{Rebulletal2011, HayesGehrkeetal2011, Souza2013, Souzaetal2014, Fitzgeraldetal2015AJ}. AstroImageJ \cite{Collinsetal2016} is a more recently developed promising free software, like SalsaJ is based on ImageJ, which is capable of multi-object differential aperture photometry for high precision light-curves over a series of images.

The AAVSO provides access for members to an online aperture photometry tool called VPhot. C-Munipack is a very useful simple free tool for photometry and matching of multiple fits images for detecting new variable stars and constructing lightcurves. Canopus, a relatively low-cost application from BdW Publishing, has been used frequently for the study of asteroid lightcurves and other photometric measurements. For those with more resources, aperture photometry can also be undertaken with the commercial packages MaximDL, Mira Pro, TheSky/CCDSoft and Prism Pro.

Aperture photometry is relatively robust but is only acceptable for relatively bright stars which are far above the sky background level and for stars that are not embedded in a crowded field of stars or other objects, such as in nebulae, compact star clusters or along the galactic plane. To overcome these issues and measure faint and/or crowded fields, Point-Spread-Function photometry needs to be undertaken. All of the common trialed and tested methods that undertake this, such as DAOPhot \cite{Stetson1987}, DoPhot \cite{Schechteretal1993} and PSFEx \cite{Bertin2011} are Linux command line tools. These tools are all potentially automated and photometry could potentially be provided to the user along with the images.

For spectroscopy, there is STSci SpecView which allows examination of optical to infrared spectra with line-identifications and sample spectra. SPLAT-VO is also a very useful spectroscopy tool containing a lot of spectra examination, measurement and manipulation tools. RSpec is a commercial spectra software available for Windows.

A broader issue underpinning these software related issues is that a lot of the pro-am software is currently written by single authors and in a closed source environment. When the single author retires from astronomy software development, then the once-indispensible software is slowly rendered defunct as the requirements of operating systems change. This is perhaps the strongest argument in supporting open source collaboration for astronomy software projects, particularly for the pro-am and education audience.

 Recent developments in web technology is on the brink of revolutionising image analysis, replacing analysis applications with analysis in web browsers. At present there are tools which would make this. A fully featured application on this is the AAVSO Vphot website\footnote{https://www.aavso.org/vphot} (previously called `Photometrica'), which allows a range of photometric analysis to be performed on images inside a web browser. It is however, only available to AAVSO members. There have been several attempts at creating Javascript libraries for read FITS data (e.g. AstroView\footnote{https://github.com/jonyardley/AstroView}, jsFITS\footnote{https://github.com/slowe/jsFITS}, fitsjs\footnote{https://github.com/astrojs/fitsjs}). A present a limitation is the size of individual FITS files which need to be uploaded and rendered in a browser, and a coordinated approach to supporting and developing such a resource.

\subsection{Lack of effective evaluation in the literature}

Much of the modern research in the field of Astronomy Education Research has largely been focussed on the improvement of non-science major ``astro101" undergraduate astronomy courses \cite{BaileyLombardi2015}.  Much evaluation has centered around the uses of concept inventories to probe student learning in large classes (e.g. \cite{Pratheretal2009}) with more of a focus on broad knowledge rather than specific learning targets \cite{Slateristar2016, Lelliott2010, Bretonesrelea2016} with less frequent studies of shifts in student motivation (e.g. \cite{Fitzgeraldetal2015RISE}) or student perceptions of the nature of science (e.g. \cite{Wallaceetal2013}).

Most robotic telescope education programs do not appear to have published peer-reviewed evaluations of their approach in the academic literature. While there are some grey literature evaluations on the internet, these tend to be unreliable, suffer from methodological flaws or significant threats to validity and be non-peer reviewed. Furthermore, there typically are only short conference papers describing the project and approach, without further details. There are some exceptions, but lack of evaluation and/or reporting is rife. This makes it hard to appraise the effectiveness of either the field in general or individual projects by themselves. For those projects that do evaluate themselves, the evaluations are particularly heterogeneous and hence direct comparisons between each project cannot be made.

For those running education projects, there are a number of guides put out by various authors that can serve as a brief introduction. ``Conducting astronomy education research: A primer" by Bailey \cite{Baileyetal2010} is a book aimed at astronomers turning to education research and evaluation for the first time. A similar guide with a more generic focus is ``Discipline-Based Education Research: A scientist's guide" by Slater \cite{Slateretal2010}. A longer, more detailed guidebook, “The 2010 User-Friendly Handbook for Project Evaluation” \cite{Frechtlingetal2010} is available for free from the NSF for those who may want to delve deeper. For those endeavouring to evaluate, it is highly recommended to find a mentor as some of the limitations, tripping hazards and complexities of this type of research are not immediately apparent to others coming from out of the field, and done well could be used for career appraisal \cite{Borrow2015}. Making the transition from science to science education is not necessarily an easy one \cite{Buxneretal2012}.

One of the simplest direct evaluation technique is the pre/post test with a concept knowledge questionnaire to see what students knew at the beginning and then be able to estimate what they learned throughout the process (e.g. \cite{Fitzgeraldetal2015RISE}) .  A sample of the most relevant concept inventories to robotic telescopes are provided in table 1. The results can also inform the instructor on what parts of the content the students are learning well or not learning from which the teaching approach can be adapted.

\begin{table}[]
\centering
\caption{Relevant concept inventories to robotic telescope projects.}
\label{my-label}
\begin{tabular}{lll}
\hline
\textbf{Name of Test}                                  & \textbf{Abbreviation} & \textbf{Citation} \\ \hline
\textit{Test of Astronomy Standards}                   & TOAST                 & Slater 2014 \cite{Slater2014}       \\ \hline
\textit{Astronomy Diagnostic Test}                     & ADT                   & Hufnagel 2000 \cite{Hufnagel2000}     \\ \hline
\textit{Star Properties Concept Inventory}             & SPCI                  & Bailey 2011 \cite{Bailey2012}       \\ \hline
\textit{Light and Spectroscopy Concept Inventory}      & LSCI                  & Bardar 2006 \cite{Bardar2006}       \\ \hline
\textit{Astronomy and Space Science Concept Inventory} & ASSCI                 & Sadler 2010 \cite{Sadler2010}      \\ \hline
\end{tabular}
\end{table}

The pre/post test approach can give the appearance of rigour and accuracy as directly from the numbers, a gain \cite{Hake1998} can be estimated and a statistical significance calculated. This is comforting to astronomers whose common research experience is largely numerical, whether observational or theoretical. However, even if the gain achieved is great and statistical significance is powerful, it still only tells the evaluator that whatever happened between the pre and the post test seems to be correlated with successful learning as measured by a multiple choice test.

It is also possible to use pre/post tests to measure other constructs such as student attitude towards science or astronomy, their philosophical views of the nature of science, their attitudes towards learning science and their opinions of their classrooms amongst many others. A very small sample of five well-known and well-used tools that could potentially be used in the evaluation of such projects (if they helped answer the right question) is presented in Table 2. There are many more. More care must be taken with such surveys as, firstly, the surveys are usually designed to measure a few fairly specific constructs and sometimes need to be administered in a particular way so a careful reading of the original paper is absolutely necessary and, secondly, shifting anything to do with a students’ psychology rather than just simple comprehension is a slow and steady process such that change is unlikely to be seen on timescales as short as a semester or a school term.

\begin{table}[]
\centering
\caption{Relevant psychometric surveys to robotic telescope projects.}
\label{my-label}
\begin{tabular}{lll}
\hline
\textbf{Name of Test}                                  & \textbf{Abbreviation} & \textbf{Citation} \\ \hline
\textit{Attitudes towards Science}             &                   & Kind 2007 \cite{Kind2007}       \\ \hline
\textit{Colorado Learning Attitudes about Science Survey}                   & CLASS                 & Adams 2006 \cite{Adams2006}       \\ \hline
\textit{Views about the Nature of Science}                     & VNOS                  & Lederman 2002 \cite{Lederman2002}     \\ \hline

\textit{Attitudes towards Astronomy}      &                   & Zeilik 1999 \cite{Zeilik1999}       \\ \hline
\textit{Constructivist Learning Environment Survey} & CLES                 & Taylor 1991 \cite{Taylor1991}      \\ \hline
\end{tabular}
\end{table}

While pre/post-testing doesn't provide a great depth of knowledge or understanding by itself, a measurement of the content knowledge gain will let the instructor know whether it worked or it did not. Hence, the pre/post test is a necessary component of project evaluation and should be undertaken by any project that can allocate the necessary time and resources, but it is not sufficient by itself to tell the full story.

Most educational research questions cannot provide a nice simple number with nice statistical error bars. Other questions that could be answered are also  limited by ethical considerations when involving human subjects in research \cite{Brogtetal2008}. Questions that can be asked in these projects such as how teacher's understandings and valuings of the scientific method are affected by research experiences \cite{Buxner2010Thesis}, what prevents teachers from undertaking inquiry in the classroom \cite{Fitzgerald2016Barriers} or how important is image ownership for robotic telescope users \cite{Gershun2014} are simply not answerable with multiple choice quantitative tests. They require qualitative exploration with the same rigour as expected in quantitative explorations.

\section{Advice when building an education program}

Finally, we present some suggestions for creating and managing education activities or programs. These are based on findings throughout this review and from our own experiences. It is not intended as a definitive guide but rather as a collection of simple-to-implement ideas which can make a substantial difference to any program.

\subsection{Goals first}
    Many projects start with a robotic telescope and then try to fit a program around it. With any educational project it is essential to outline the goals of your project before considering what tools you need to achieve those goals. You will not only be able to refer to these goals when you are creating the educational materials, software, and other  items for a successful program. It will help your program to stay focused, making it easier to achieve and support.  If your initial goals are successful or even too simplistic, you can always revise them after trialing your program. The evaluation of your program should reference these goals and how well they were achieved by the audience.

\subsection{Who is the audience?}
    You might think your program will appeal to everyone, and while the essence of the program might have broad appeal, identifying the key group or groups who will (or should) benefit most from your program is very useful. It will guide your choice of activities and help your program to remain focused. In doing so, try not to overclaim and say that your participants will or can do things which are beyond the scope of the program.

\subsection{Does your project match the instrumentation?}
    Some telescopes are too big to image the moon, the planets or bright stars. Some telescopes are too small or track too poorly to image distant galaxy clusters and faint planetary nebula. Some telescope networks are not set up to easily periodically observe a variable star or asteroid. Some telescopes are not in optimal locations and can easily be rained out hampering attempts at direct control. Some telescopes do not have science grade filters available to do research-grade photometry. Most telescopes (so far) do not have spectroscopes. All telescopes are different and matching your intended observations to the capacity of the instrumentation or user interface is very important.

\subsection{Simple user interface}
    If you run a robotic telescope for education your user interface will be the key component to the success of your program. Your choice of goals and understanding of who your audience is will be critical in your choice of interface. There are many off-the-shelf interface solutions but almost all of them assume a working knowledge of aspects of astronomy that professionals and amateurs take for granted (e.g. catalogue names for astronomical objects are relatively unknown to teachers, a telescope's field of view will be a mystifying concept to students). One of your goals may be educating your audience to a sufficient level to use the interface. Be wary of requiring the audience to acquire skills and knowledge which are too specific to your program. It may work better for your program to use a simplistic interface which encourages a larger audience to participate.

\subsection{A document is never as good as an expert}
    Do not assume any of your users will read documentation. Only the keenest of your audience will hunt for information if it is not immediately obvious. Do not assume that teachers who have been trained in the activities of your program will remember the full details. If you can provide in-person mentoring to your audience, it will dramatically improve not only their understanding of the program but also their investment in it. You might also consider running regular webcasts/webinars. These should involve a way for the audience to interact (e.g. ask questions) and be archived for your audience to watch at their convenience.

\subsection{Time is precious and short}
    People new to robotic telescopes and astronomical data can take a long time to understand the various concepts, tools and tasks. Education users, whether in a formal classroom or an informal outreach situation, are typically very time limited. Unless your project is focussed on mentoring users through long projects, it is best to remove any approach or material that unnecessarily uses up the limited time of the user.

\subsection{Sustaining resources are necessary}
    As noted in previous publications, e.g.  \cite{FitzgeraldHollowEtAl2014}, one of the major failing points of any education project is a lack of funding. Sometimes very well meaning projects come into existence on the back of a successful traditional 3-year grant, only to go through a lifecycle of necessary development and to disappear as soon as the grant funding dries up. Sufficient thought about the longevity of resources needs to be taken into account in any approach as education, nearly by definition, is something that occurs over a longer, more sustained period. The failing point in most projects is usually due to lack of monetary resources, but resources such as the voluntary time of skilled people as well as in-kind telescope time provision must also be taken into account.

\subsection{Good Design and Continual Improvement}
    The first version of any materials or approach used in any project will not be perfect. It will likely have errors and will not quite fit the intended audience. This is inevitable, try to use known principles of instructional design and how people learn when creating your initial attempts. Where possible, observe your activity being used or delivered to the target audience, and see how it differs from your intended implementation and why. Some differences may be very obvious while some may (or can) only appear after analysing your evaluation results. In light of this information, revisit and reform your provided materials and approach. An ethos of continual evaluation and revision will improve the quality of your project over time.

\section{Conclusion}

In this review, we have provided the first systematic overview of the role of optical robotic telescopes in education. The combination of this remote observing technology with education has shown much promise and has inspired much effort from many people, over a period of decades. Unfortunately, it largely suffers from the same problem that many other education technologies have; only a small fraction of incredibly keen users have included it in their regular programs but it is still waiting to be accepted by the majority of educators.

As noted by many authors cited throughout this review paper, careful and considered project design is a necessity for any of these educational projects to achieve success. Due to the significant lack of evaluation of most of these projects, what qualifies as relative success in an objective manner is hard to define. To help the field of robotic telescope education develop, it is incredibly important that more research is undertaken in the domain overlapping robotic telescopes and education. In particular, project evaluations using well-tested concept inventories, such as those in table 1, and deep qualitative investigations, matched with detailed and thorough descriptions of theoretical framework and project approach will help this field develop effectively. In doing so, it is hoped that this will help boost our educational capacity to bring it more in line with the current technical capacity.

There is however much to be optimistic about. Of the known effective evaluations, project design descriptions and even anecdotal successes, many of these have been concentrated into the last five or so years. The burgeoning technical capacity from the variety of telescope time providers is facilitating ever increasing ease of access for diverse audiences. Some of the educational projects are now approaching, or are currently in, their second decade of operation giving some hope for the stability of the field.

We conclude this review by noting that technical capacity, while still in development, is currently outstripping educational capacity. Astronomy education would significantly benefit from simple, high-level, well thought-out and innovative approaches to using robotic telescopes.


\begin{thebibliography}{99}
\bibitem {Graham2015}
Graham, GG How is success defined and measured in online citizen science: a case study of Zooniverse projects. Computing in science and engineering. 23p.
\end{thebibliography}
\end{document}